\documentclass[pdftex, twocolumn, epjc3]{svjour3}
\pdfoutput=1

\usepackage{placeins}
\usepackage{multirow}
\usepackage[utf8]{inputenc}
\usepackage{color}

\RequirePackage[detect-all=true,group-digits=true,group-separator={,},binary-units=true]{siunitx}
\RequirePackage{xspace}
\RequirePackage{graphicx}
\usepackage{lineno}

\def\bracketbar{\hbox{\kern-9pt\raise1pt%
    \hbox{{\tiny(}{\lower1.5pt\hbox{\bf--}}{\tiny)}}}}

\journalname{Eur. Phys. J. C}
\RequirePackage{float}
\RequirePackage{mathtools}
\RequirePackage{color,soul}
\RequirePackage{xcolor}
\usepackage{mathtools}  
\usepackage{lineno}

\RequirePackage{latexsym}
\RequirePackage[numbers,sort&compress]{natbib}
\RequirePackage[colorlinks,citecolor=blue,urlcolor=blue,linkcolor=blue]{hyperref}
\usepackage{multirow}
\usepackage{multicol} 
\usepackage[english]{babel} 
\usepackage{microtype}
\usepackage{comment}
\usepackage[utf8]{inputenc}
\usepackage{dcolumn}
\usepackage{bm}

\begin{document}
\title{Scintillation light detection in the 6-m drift-length ProtoDUNE Dual Phase liquid argon TPC
}

\date{\today}

\authorrunning{DUNE Collaboration}
%

\author{The DUNE Collaboration\\\\
       A.~Abed~Abud\thanksref{Liverpool,CERN}
       \and B.~Abi\thanksref{Oxford}
       \and R.~Acciarri\thanksref{Fermi}
       \and M.~A.~Acero\thanksref{Atlantico}
       \and M.~R.~Adames\thanksref{Tecnologica }
       \and G.~Adamov\thanksref{Georgian}
       \and M.~Adamowski\thanksref{Fermi}
       \and D.~Adams\thanksref{Brookhaven}
       \and M.~Adinolfi\thanksref{Bristol}
       \and A.~Aduszkiewicz\thanksref{Houston}
% ...
}

\institute{
% ...
\pagebreak[0] Universidad del Atl\'antico, Barranquilla, Atl\'antico, Colombia\label{Atlantico}
% ...
\and\pagebreak[0] University of Bristol, Bristol BS8 1TL, United Kingdom\label{Bristol}
        \and\pagebreak[0] Brookhaven National Laboratory, Upton, NY 11973, USA\label{Brookhaven}
% ...
\and\pagebreak[0] CERN, The European Organization for Nuclear Research, 1211 Meyrin, Switzerland\label{CERN}
% ...
\and\pagebreak[0] Fermi National Accelerator Laboratory, Batavia, IL 60510, USA\label{Fermi}
% ...
\and\pagebreak[0] Georgian Technical University, Tbilisi, Georgia\label{Georgian}
% ...
\and\pagebreak[0] University of Houston, Houston, TX 77204, USA\label{Houston}
% ...
\and\pagebreak[0] University of Liverpool, L69 7ZE, Liverpool, United Kingdom\label{Liverpool}
% ...
\and\pagebreak[0] University of Oxford, Oxford, OX1 3RH, United Kingdom\label{Oxford}
% ...
\and\pagebreak[0] Universidade Tecnol\'ogica Federal do Paran\'a, Curitiba, Brazil\label{Tecnologica }
% ...
}

\onecolumn
\maketitle
\twocolumn
\sloppy

\begin{abstract}
DUNE is a dual-site experiment for long-baseline neutrino oscillation studies, neutrino astrophysics and nucleon decay searches. ProtoDUNE Dual Phase (DP) is a 6\,$\times$\,6\,$\times$\,6\,m$^3$ liquid argon time-projection-chamber (LArTPC) that recorded cosmic-muon data at the CERN Neutrino Platform in 2019-2020 as a prototype of the DUNE Far Detector. Charged particles propagating through the LArTPC produce ionization and scintillation light. The scintillation light signal in these detectors can provide the trigger for non-beam events. In addition, it adds precise timing capabilities and improves the calorimetry measurements. In ProtoDUNE-DP, scintillation and electroluminescence light produced by cosmic muons in the LArTPC is collected by photomultiplier tubes placed up to 7\,m away from the ionizing track. In this paper, the ProtoDUNE-DP photon detection system performance is evaluated with a particular focus on the different wavelength shifters, such as PEN and TPB, and the use of Xe-doped LAr, considering its future use in giant LArTPCs. The scintillation light production and propagation processes are analyzed and a comparison of simulation to data is performed, improving understanding of the liquid argon properties.  


\end{abstract}
\section{Introduction}
\label{sec:intro}

The Deep Underground Neutrino Experiment (DUNE)~\cite{DUNEtdrv1} aims to address key questions in neutrino physics such as measuring the CP violating phase and the neutrino mass ordering with an intense muon neutrino beam produced at Fermilab~\cite{DUNE_LBL}. The physics program also addresses non-beam physics such as nucleon decay and beyond the Standard Model searches~\cite{DUNE_BSM} and the detection and measurement of the electron neutrino flux from a potential core-collapse supernova within our galaxy~\cite{DUNE_SNB}. DUNE will consist of a near detector placed at Fermilab close to the production point of the neutrino beam of the Long-Baseline Neutrino Facility (LBNF) to measure the unoscillated neutrino interaction rate, and four 17-kt liquid-argon time-projection chambers (LArTPCs) as far detector in the Sanford Underground Research Facility (SURF) at 4300\,m.w.e. depth at 1300\,km from Fermilab~\cite{DUNEtdrv4,duneIDRv3} where the neutrino interaction will be measured after neutrinos have oscillated. 

\begin{figure*}[ht]
    \centering
    \includegraphics[width=\textwidth]{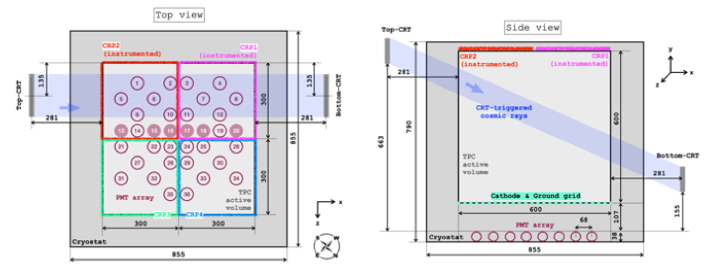}   
    \caption{Views of ProtoDUNE-DP. Dimensions and positions of the major elements are indicated (in units of centimeters). The drift direction corresponds to the $y$-axis. The PMTs are represented with circles and, in the top view, the empty circles correspond to PEN PMTs and the filled circles to TPB PMTs.}
    \label{fig:diagram}
\end{figure*}

The ProtoDUNE Dual Phase (DP) detector~\cite{wa105,Cuesta:2019yeh} was operated from 2019 to 2020 at the CERN Neutrino Platform to demonstrate the LArTPC DP technology at large scale as a prototype for one of the DUNE far detector modules. ProtoDUNE-DP has an active volume of 6$\times$6$\times$6 m$^{3}$ corresponding to an active mass of 300\,t (total LAr mass of 750\,t), being the largest DP LArTPC ever operated. In ProtoDUNE-DP the electric drift field is oriented in the vertical direction, causing the electrons to drift vertically towards the anode at the top. The ionization charge is then extracted, amplified, and detected in gaseous argon above the liquid surface by the charge readout planes (CRPs). The DP technology allows a good signal to noise ratio ($>$10 for 6-m drift at 500\,V/cm drift field) and a fine spatial resolution ($\sim3.125\times3.125\times0.64\,mm^3$)~\cite{Aimard_2021} enabling the construction of large active volumes making efficient use of the LAr volume. The CRP consists of the extraction grid which bounds the active volume, large electron multipliers (LEMs) to produce the charge avalanche, and the anode to collect the electrons. In ProtoDUNE-DP, two fully instrumented CRPs of 3\,$\times$\,3\,m$^2$, and two CRPs without LEMs are installed. The scintillation light signal is collected by a photon detection system (PDS) constructed out of photo-multiplier tubes (PMTs). The PDS goals are to provide a trigger and to determine precisely the event time, with capability to perform calorimetric measurements and particle identification. Two Cosmic Ray Tagger (CRT) panels with eight scintillator bars (1.44\,m\,$\times$\,0.12\,m) per panel are placed on opposite sides of the ProtoDUNE-DP cryostat to trigger on muon-tracks passing through both CRTs. Fig.~\ref{fig:diagram} shows a diagram of ProtoDUNE-DP with the layout and dimensions of the CRPs, PMTs and CRTs. The PDS layout is optimized to maximize the collected light~\cite{anne_phd}.

As charged particles pass through LAr, they create pairs of positively charged argon ions (Ar$^+$) and free electrons and also produce excited atoms (Ar$^{\ast}$). When ionized and excited argon atoms couple to neutral Ar atoms, they produce the molecular states Ar$^+_2$ and Ar$^{\ast}_2$, respectively. The first one eventually recombines with an electron producing Ar$^{\ast}_2$. In both processes (recombination and excitation) the decay of the final state results in the emission of a vacuum ultraviolet (VUV) photon within a wavelength centered at $127\pm8$\,nm~\cite{Heindl:2010zz} constituting the primary scintillation light (S1) signal: 

\begin{eqnarray}
\mathrm{Rec.}:& \; \mathrm{Ar}^+ + \mathrm{Ar} \rightarrow \mathrm{Ar}_2^+ + \mathrm{e}^- \rightarrow \mathrm{Ar}_2^* \rightarrow 2\mathrm{Ar} + \gamma,  \\
\mathrm{Exc.}:& \; \mathrm{Ar}^* + \mathrm{Ar} \rightarrow \mathrm{Ar}_2^* \rightarrow 2 \mathrm{Ar} + \gamma.
\end{eqnarray}

In order for the recombination process to occur, an electron cloud surrounding the Ar$^+_2$ is needed. Hence the scintillation light yield depends on the electric field. 
Ar$^{\ast}_2$ has two possible states, a singlet and a triplet state. The singlet transition to the ground state Ar$_2$ has a short decay time $\tau_\mathrm{fast}\sim$6\,ns, while the triplet transition to the same ground state is allowed only because of spin-orbit coupling and has a much longer lifetime $\tau_\mathrm{slow}\sim$1.6\,$\mu$s~\cite{Hitachi}. 

In addition, electroluminescence secondary scintillation light, called S2, is produced in the gas phase of the dual phase LArTPC when electrons, extracted from the liquid, are accelerated in the electric field of the LEMs. The S2 signal also has 127\,nm wavelength. The time difference between the S1 and the S2 signals reflects the drift time of the original ionization in the liquid phase up to the gas phase, and the S2 duration, which can be up to hundreds of microseconds, is directly related to the track topology and covers the span of the electron drift time. 

\begin{figure}[ht]
    \centering
    \includegraphics[width=0.5\textwidth]{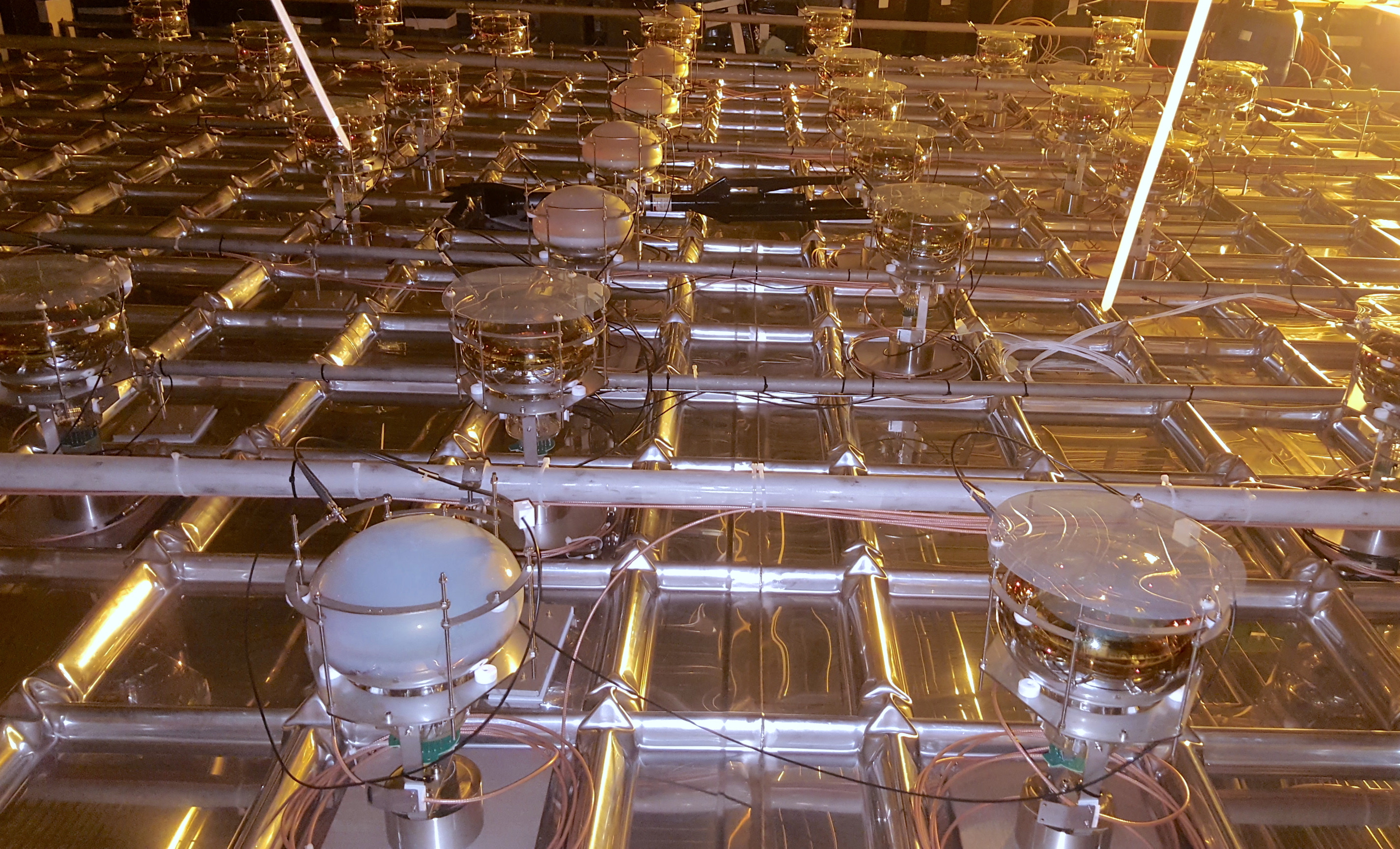}
    \caption{ProtoDUNE-DP PDS formed of 36 PMTs installed at the bottom of the cryostat at CERN. The front left PMT has TPB coating and the front right PMT has PEN sheet.} 
    \label{fig:pds}
\end{figure}

The PDS of ProtoDUNE-DP~\cite{protoDUNElight} is formed of 36 8-inch cryogenic R5912-02MOD PMTs from Hamamatsu~\cite{protoDUNEPMTs, Belver:2020qmf}, placed below the cathode grid. As the PMTs are not sensitive to VUV light, a wavelength shifter converts 127-nm photons into visible photons. Two different wavelength shifters were deployed. A sheet of polyethylene naphthalate (PEN) is placed on top of 30 PMTs and the other 6 PMTs have tetraphenyl butadiene (TPB) directly coated on them. Fig.~\ref{fig:pds} shows a picture of the PDS installed in ProtoDUNE-DP.

A light calibration system (LCS) was developed for ProtoDUNE-DP to monitor the PMT performance and obtain an equalized PMT response~\cite{Belver:2019lqm}. The light source consists of 6 blue LEDs and a silicon photo-multiplier (SiPM) as reference sensor. The LEDs use a Kapustinsky~\cite{1985NIMPA.241..612K} circuit as LED driver, and have a wavelength of 465\,nm that matches the PMT maximum quantum efficiency. The calibration light is transmitted through a fiber system with a fiber-end pointing at each PMT photocathode.

This paper describes the ProtoDUNE-DP PDS performance after 18 months of operation considering its future use in giant LArTPCs with a particular focus on the wavelength shifting techniques. A study of the LAr scintillation and electroluminescence light production and propagation over large distances is presented. In section~\ref{sec:data}, the ProtoDUNE-DP operation and collected light data are described. Section~\ref{sec:per} details the performance of ProtoDUNE-DP PDS. In section~\ref{sec:sim} the light simulation tools are detailed. The analyses of scintillation light propagation and production are discussed in section~\ref{sec:light} and the cosmic muon rate measured in ProtoDUNE-DP is described in section~\ref{sec:muon}. Studies using the electroluminescence light signal are summarized in section~\ref{sec:s2}. Finally the performance of the system using Xe-doped LAr is reviewed in section~\ref{sec:xe}.
\section{ProtoDUNE-DP PDS operation at CERN}
\label{sec:data}

ProtoDUNE-DP collected cosmic-ray data from June 2019 until November 2020, operating with pure LAr and Xe-doped LAr in different conditions of electric fields. Muons are the most abundant charged particles in cosmic rays at surface, together with protons ($\sim$1\%), and electrons and positrons ($\sim$0.1\%). No beam data were taken with ProtoDUNE-DP. 

\begin{figure}[ht]
    \centering
    \includegraphics[width=0.45\textwidth]{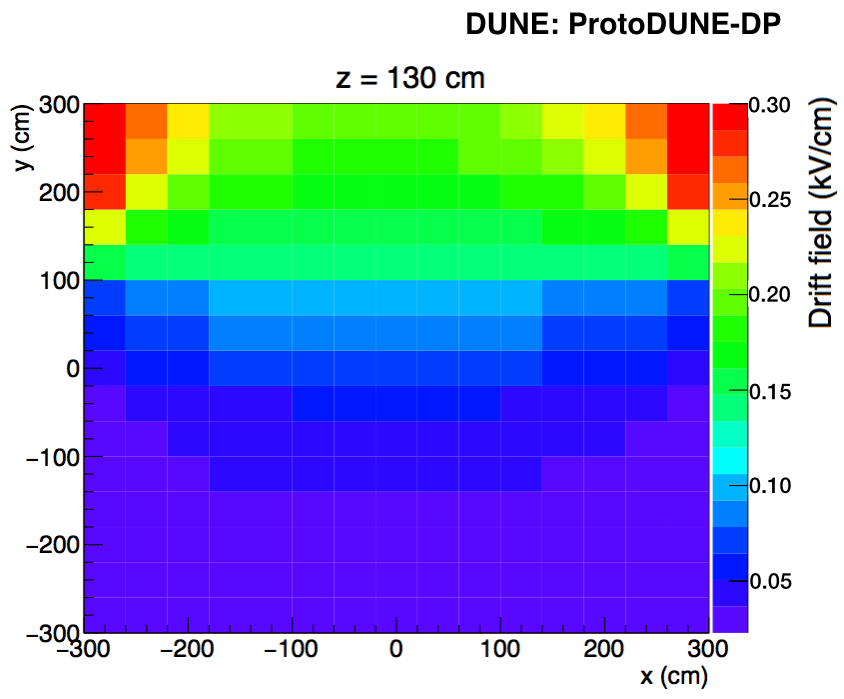}
    \caption{Map of the drift field in ProtoDUNE-DP with cathode at -50 kV simulated using COMSOL. A particular plane of the active volume is shown (vertical plane at 130\,cm from a field cage wall). The drift direction is along the y-axis and the color scale represents the electric field strength. The discontinuity at y$\sim$100\,cm is due to the HV failure described in the text.}
    \label{fig:mapfield}
\end{figure}

The ProtoDUNE-DP operation faced issues that impacted the electric field conditions. A short circuit  between the high-voltage (HV) source and a field-cage ring at approximately 1/3 the field-cage depth  prevented reaching the nominal cathode voltage (-300 kV). The cathode HV was set at -50\,kV and the drift field in the operation conditions was fairly uniform in the top part of the drift and away from the field cage walls. A COMSOL~\cite{COMSOL} simulation of the resulting drift field is shown in Fig.~\ref{fig:mapfield}. Additionally, CRP operation was not straightforward due to the presence of bubbles on the LAr surface causing LEM and grid discharges. Given these limitations, data taken with the CRPs in combination with the PDS were very limited and the data presented in this paper were taken with the PDS alone. The operation conditions of ProtoDUNE-DP were 88\,K with a thermal gradient present in the detector less than 20\,mK and 1045\,mbar.

The PDS started its operation as soon as the detector was purged with argon gas in June 2019. ProtoDUNE-DP operated fully filled with LAr from August 2019 until May 2020. In June 2020, an intervention on the HV extender was carried out with the aim of fixing the short circuit although the issue was not solved. In July 2020, the detector was re-filled using $\sim$230 ton of Xe-doped LAr from ProtoDUNE-SP~\cite{protoDUNESP} contaminated with N$_2$. In August 2020, operations were resumed and two additional N$_2$ injections took place to measure the effect of N$_2$ contamination in the light attenuation length. Table~\ref{tab:XenonData} summarizes the different doping concentrations in ProtoDUNE-DP.

\begin{table}[!ht]
\begin{center}
\caption{Summary of Xe and N$_2$ doping concentrations in ProtoDUNE-DP. Units are ppm in mass for Xe, and ppm in volume for N$_2$.}
\label{tab:XenonData}
\begin{tabular}{c c c}
\hline
    Situation & [Xe] (ppmm) & [N$_2$] (ppmv) \\
    \hline
    LAr + Xe + N$_{2}$ & 5.8 &  2.4 \\
    1$^{st}$ N$_{2}$ injection & 5.8 & 3.4\\
    2$^{nd}$ N$_{2}$ injection & 5.8 & 5.3 \\
    \hline
\end{tabular}

\end{center}
\end{table}

The PDS took data on a daily basis during short time periods (typically 1-2 hours/day). PMT HVs had to be switched off during the operation of some monitoring systems that emit light, like the purity monitors or the cryogenic cameras, to avoid any damage to the PMTs. Especially at the beginning of the data taking period, this happened very often to survey the status of the liquid surface and charge readout planes. Longer time periods of PDS data taking ($\sim$12 hours) were allowed, typically at night, proving the stability of the PDS. 

\begin{figure*}[ht]
    \centering
    \includegraphics[width=0.65\textwidth]{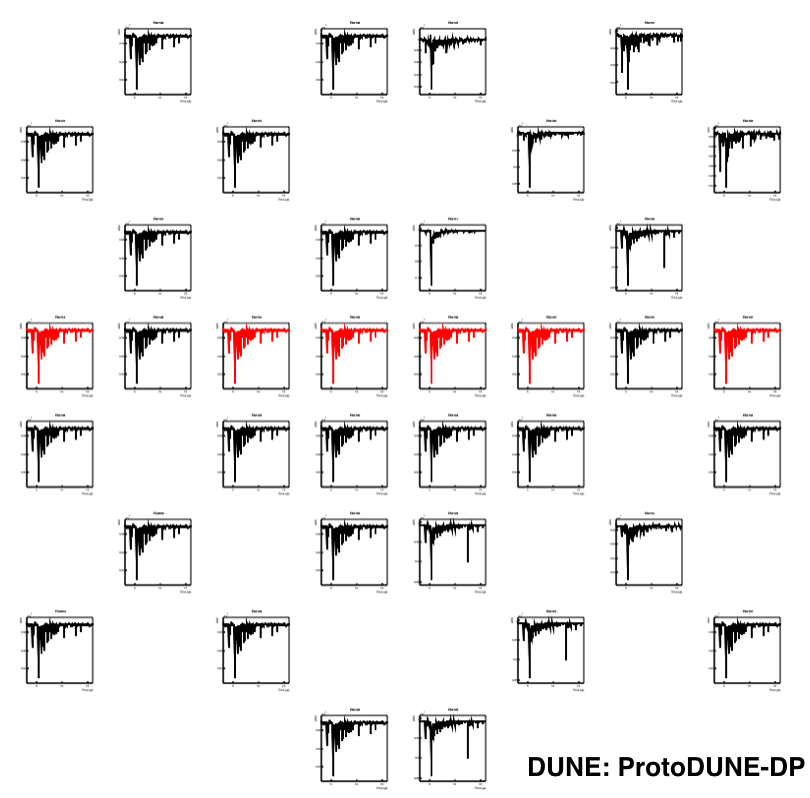}
    \caption{Example of a PMT self-trigger event showing the waveforms corresponding to the 36 PMTs according to their relative position in the detector. The PEN PMTs are shown in black and TPB PMTs in red. The $x$-axis range is 0-16\,$\mu$s for all PMTs while the $y$-axis range (in ADC counts) varies for each PMT and is optimized to best display the S1 signal.}
    \label{fig:EventExample}
\end{figure*}

A dedicated light data acquisition and calibration software was developed for ProtoDUNE-DP~\cite{ProtoDUNEDP_LACS}. The software allows the user to choose the acquisition  trigger mode, control and define the acquisition settings (front-end and high-voltage), and provides the graphical user interface. The light readout front-end electronics is based on the commercial ADC V1740 from CAEN \cite{CAEN}. This 12-bit VME digitizer has 64 analog input channels with 2\,Vpp dynamic range and a maximum sampling rate of 62.5\,MS/s. All data presented in this paper were taken with 16-ns sampling, and different time acquisition windows depending on the configuration (from 2\,$\mu$s to 4\,ms). 

\begin{figure}[ht]
    \centering
    \includegraphics[width=0.35\textwidth]{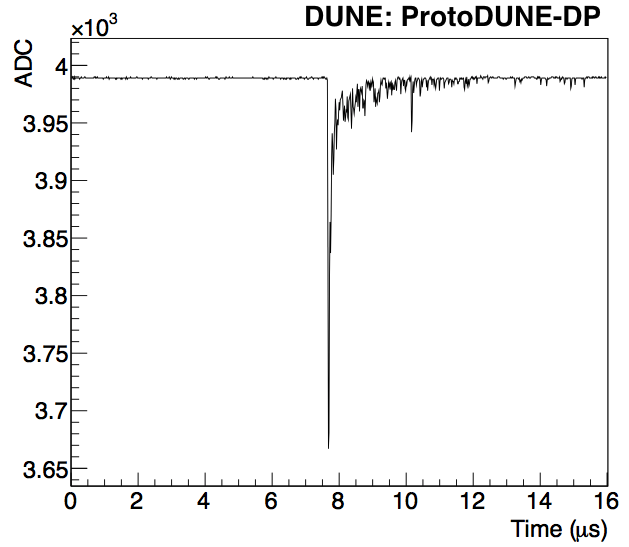}
    \caption{Example of a scintillation light event in a ProtoDUNE-DP PMT at a gain of $10^7$ in LAr without electric field.}
    \label{fig:WF}
\end{figure}

Individual PMT waveforms were recorded for analysis, so each event contains 36 waveforms. An example of a PMT self-trigger event is shown in Fig.~\ref{fig:EventExample} and an individual PMT waveform is shown in Fig.~\ref{fig:WF}.  In the off-line analysis, various properties of the waveform are evaluated: the integrated charge in a given time period (in photo-electron units, PE), the time when the waveform reaches the minimum value ($t_0$), the amplitude of the event, and the baseline mean and standard deviation (STD). 

The PDS system took data in three trigger modes:
\begin{itemize}
\item PMT self-trigger: Provided by the PDS when a set number of PMT signals (or more) over a given threshold are in coincidence. The trigger rate is in the Hz-kHz range depending on the customizable threshold and the number of PMTs required to pass the threshold. In this trigger mode, not only single tracks are collected, but also showers and multi-track.
\item CRT trigger: In this case the PDS receives an external
trigger signal from the CRT planes at an average rate of 0.3 Hz. The CRT trigger mode allows the recording of single tracks with known topology.
\item Calibration mode: An external trigger signal is received from the light calibration system at 1\,kHz synchronized with the calibration light pulse sent to the PMTs. This mode also allows taking data with random trigger at a configurable rate turning off the LEDs.
\end{itemize}

A total of 130.7 million events were acquired during a live time of 675 hours. Table~\ref{tab:DataSummary1} shows the summary of the data taken for the different trigger configurations. As shown in Table~\ref{tab:DataSummary2}, most of the light data were acquired without electric drift field, and therefore contain only primary scintillation signals. Events with secondary light signals were collected for different drift and amplification field conditions. An important sample of events was recorded during the commissioning phase of the system with varying conditions (test mode). 

\begin{table}[!ht]
\begin{center}
\caption{Summary of trigger conditions indicating the number of events and total time of data taken in the different trigger configurations.}
\label{tab:DataSummary1}
\begin{tabular}{ccc}
\hline 
Trigger & \# of events (M) & Time (h) \\ 
\hline 
PMT  & 85.3 & 96 \\
CRT Panels & 0.6 & 515 \\
Calibration & 30 & 42 \\
Random  & 14.7 & 21 \\
\hline 
Total   & 130.7 & 675 \\
\hline 
\end{tabular} 
\end{center}
\end{table}

\begin{table}[!ht]
\begin{center}
\caption{Summary of the detector configurations indicating the number of events and total time of data taken with different voltage across LEMs.}
\label{tab:DataSummary2}
\begin{tabular}{cccc}
\hline 
Drift field & LEMs voltage  & \# of events (M) & Time (h) \\ 
\hline 
OFF  & -       &  85.1 & 342\\
\hline 
ON
  & 0\,kV   & 13.6 & 72\\
  & 2.5-3.6\,kV   & 7.7 & 212\\
\hline 
Test & -  & 23.2 & 48\\
\hline 
Total & -  & 130.7 & 675\\
\hline 
\end{tabular} 
\end{center}
\end{table}


\section{ProtoDUNE-DP PDS performance}
\label{sec:per}


All 36 PMTs were operational throughout of data taking, allowing the validation of the basic performance of the PDS. A coincident primary scintillation light signal (S1) detected by all 36 PMTs is visible in Fig.~\ref{fig:EventExample}. The time alignment among PMT signals has been measured for all channels to be better than 16\,ns.

\begin{figure}[ht]
    \centering
     \includegraphics[width=0.45\textwidth]{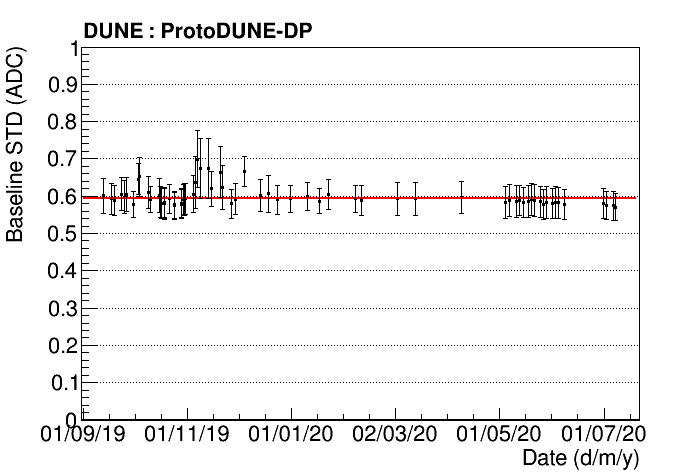}
    \caption{Baseline mean STD during ProtoDUNE-DP operation. A fit to the mean value ($0.6\pm0.1$\,ADC) is shown in red and error bars represent the differences among the different PMTs.}
    \label{fig:baseline}
\end{figure}

The low noise in the baseline of the signals is notable, as the baseline presents very small fluctuation: $0.6\pm0.1$\,ADC ($0.29\pm0.05$\,mV). The error includes the differences among PMTs and the stability along the time. In addition, the baseline stability with time is shown in  Fig.~\ref{fig:baseline}.

With the aim of understanding the performance of the PDS, the gain calibration results are shown in section~\ref{sec:cal}, the single photo-electron (SPE) is characterized in terms of amplitude and rate in section~\ref{sec:spe}, the relative performance of PEN and TPB is evaluated in section~\ref{sec:wls}, and the scintillation time profile is studied in section~\ref{sec:time}. 

\subsection{Calibration} 
\label{sec:cal}

\begin{figure*}[ht]
  \centering
  \includegraphics[width=0.95\textwidth]{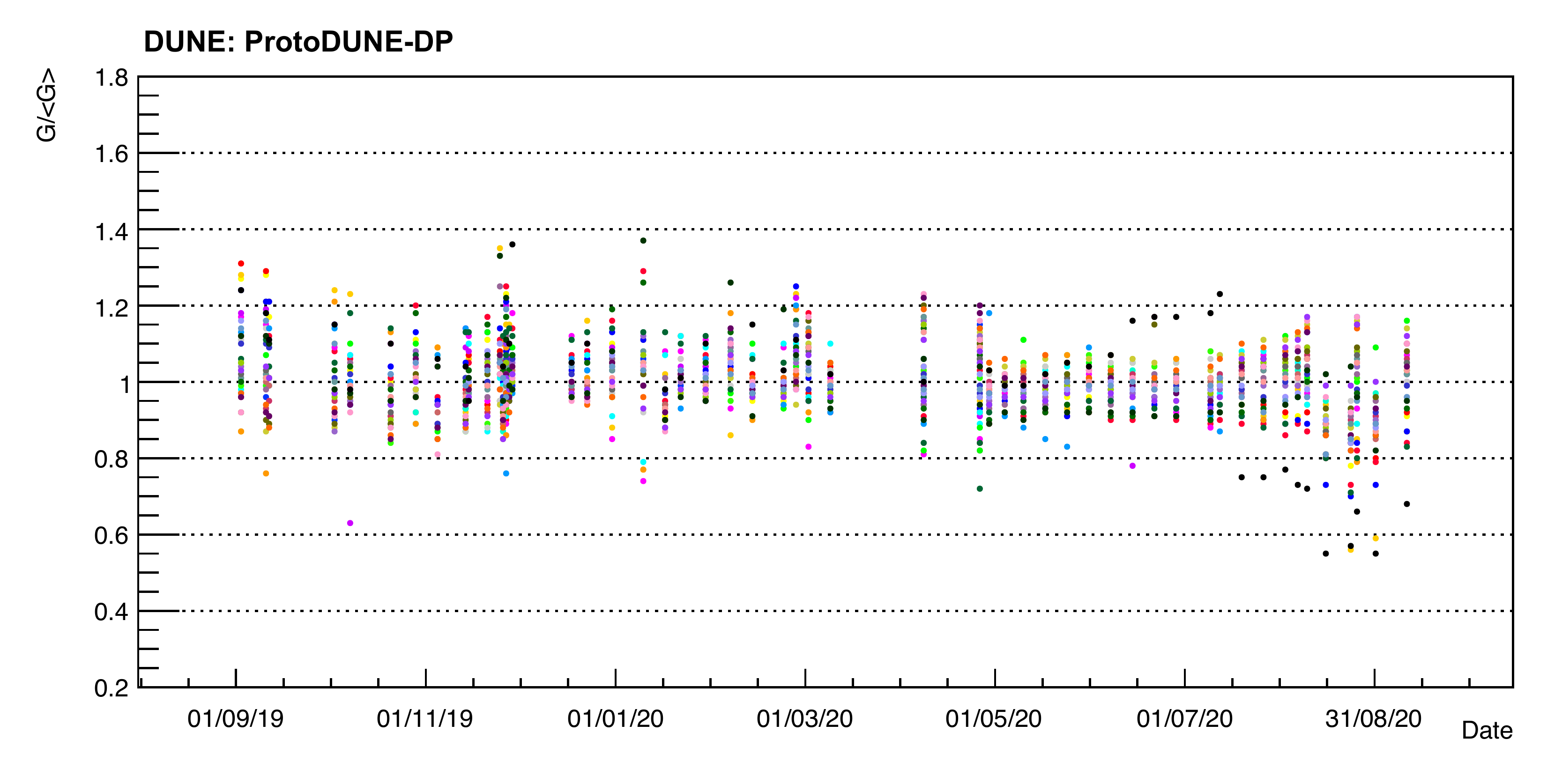}
  \caption{Gain ($G$) stability with time for the 36 PMTs (one color per PMT) evaluated at 1500\,V represented with respect to the average gain $<G>$.}
   \label{fig:Gstab}
\end{figure*}

The main goal of the LCS is to calibrate the PMT response by determining the PMT gain during the operation of the detector. It is important to guarantee equalized PMT response and to measure the light collected in PE units. An accurate measurement of the collected light is essential for calorimetry and to estimate the detection efficiency of the PDS. The PMTs are typically operated at a gain in the range from $1\cdot10^7$ to $5\cdot10^7$ as a compromise between maximizing the sensitivity to the SPE and minimizing ADC-saturated events. The LCS illuminates the PMT photocathode at the SPE level in order to determine the PMT gain. The calibration light rate is kept to $\sim$kHz to avoid PMT fatigue~\cite{Belver:2019lqm}.

The gain calibration method, based on measuring the SPE charge at a given voltage, is described in~\cite{Belver:2019lqm}. During operation, PMTs were biased at the HV required to achieve the target gain according to the calibration results. Calibrations were carried out weekly and a gain correction based on the closest calibration in time is applied in the analysis.

The calibrations performed during the detector operation allow to monitor the PMT gain stability with time. Fig.~\ref{fig:Gstab} presents the gain evolution of all the PMTs during one year of regular operation of the LCS. It should be noted that the PMTs were switched on and off every day (sometimes several times on the same day). Despite this, it can be seen that the PMT gains are quite stable with time. In particular, the average value of the gain STD at 1500\,V for 36 PMTs is $9\%$.

\subsection{Single photo-electron characterization} 
\label{sec:spe}

The capability of measuring low energy signals depends on the signal-to-noise ratio and on the SPE rate from various background sources. 

\begin{figure}[ht]
    \centering
    \includegraphics[width=0.45\textwidth]{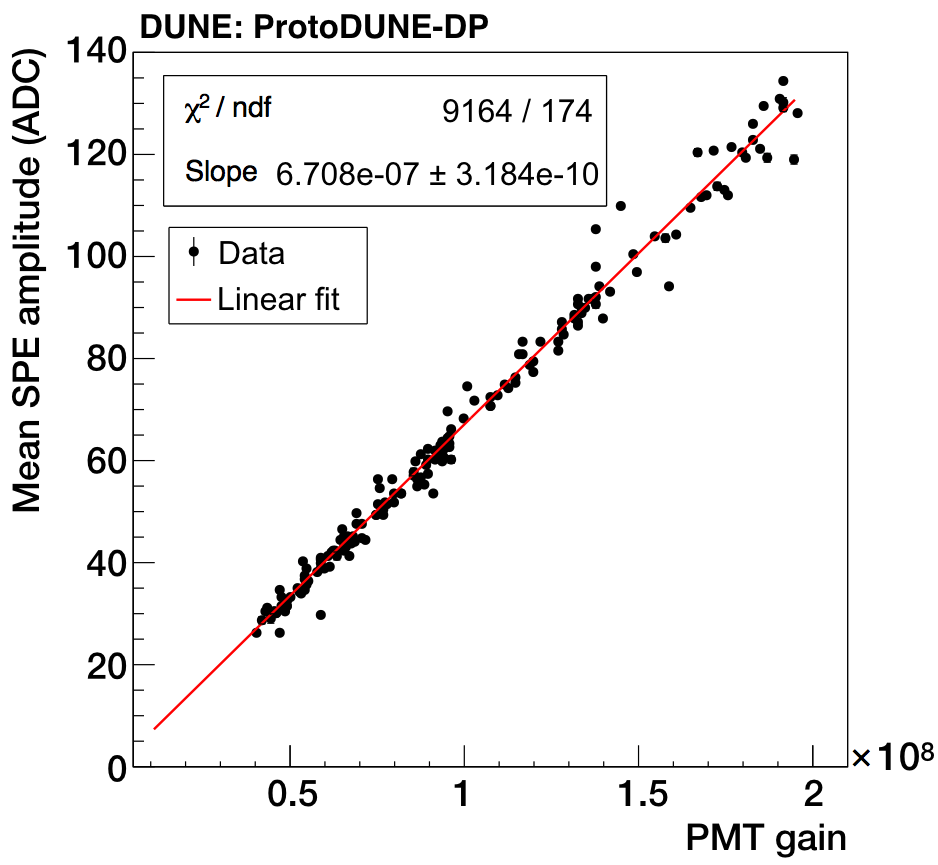}
    \caption{Mean SPE amplitude from Gaussian fits as a function of the PMT gain, and linear fit. All PMTs are included but only data corresponding to gains that allow a proper fit are considered.}
    \label{fig:SPEamplitude}
\end{figure}

\begin{figure}[ht]
   \centering
   \includegraphics[width=0.35\textwidth]{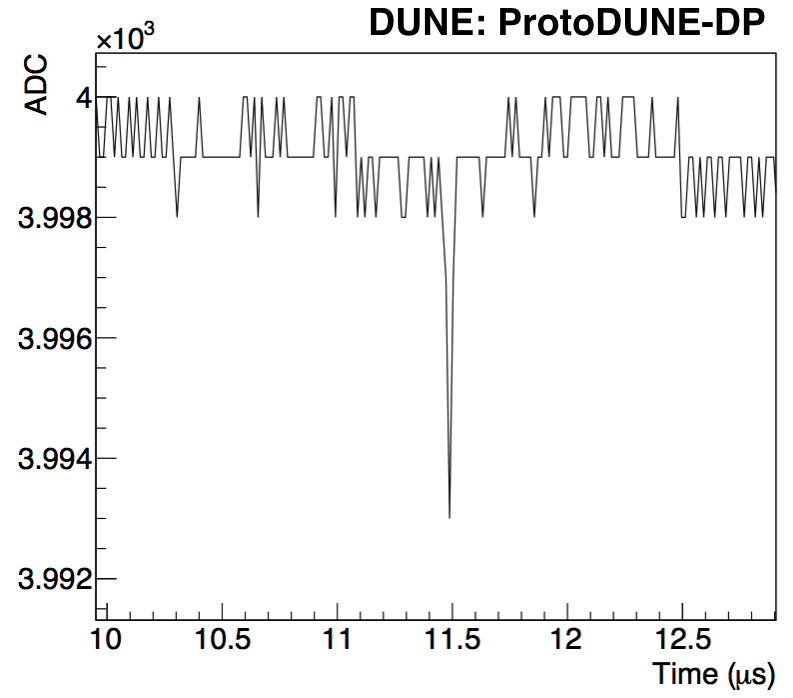}
   \caption{Example of SPE pulse in a PMT waveform ($10^7$ gain).}
   \label{fig:SPE}
\end{figure}

The SPE amplitude is characterized as a function of the PMT gain using calibration data. The SPE amplitude spectra are fitted with a Gaussian function for each PMT in order to obtain the mean amplitude ($\mu$) and its corresponding standard deviation ($\sigma$) as functions of the PMT gain. The correlation between mean amplitude and gain is shown in Fig.~\ref{fig:SPEamplitude}, which allows the average SPE amplitude at any gain to be extracted. For instance, at a gain of $10^7$, the SPE amplitude is 7\,ADC counts, implying a signal-to-noise ratio greater than 11 thanks to the small fluctuation of the baseline in the PMT waveforms, while $\sigma$ is 2\,ADC counts. An example of a SPE pulse in a PMT waveform at a gain of $10^7$ is shown in Fig.~\ref{fig:SPE}. The SPE pulses have a full width at half maximum (FWHM) of 1 time sample (59\%, 16 ns) or 2 time samples (40\%, 32 ns).

The SPE rate is computed by counting the SPE pulses in the data samples taken with the random trigger. Pulses with an amplitude within the expected range ($\mu\pm2\sigma$) are considered, ignoring the parts of the waveform affected by S1 signals larger than 1\,PE. Without drift field, a SPE rate of $\sim$350\,kHz is obtained for TPB PMTs, whereas PEN PMTs are less sensitive and a rate of $\sim$170\,kHz is measured. 

There are several light contributions at the SPE level to the low-energy background  detected  by  the  PMTs. First, according to data analysis, cosmic muons are expected to be detected as SPE at $\sim$35\,kHz. Second, according to a dedicated simulation, natural radioactivity, mainly from 1.01\,Bq/kq $^{39}$Ar~\cite{Benetti:2006az}, 115\,mBq/kg $^{85}$Kr~\cite{Benetti:2006az} and 0.09\,mBq/kg $^{42}$Ar~\cite{Barabash:2016lru}, is expected to contribute with $\sim$20\,kHz. Lastly, it is known that the PMT dark current contributes with a SPE rate of $\sim$1.7\,kHz~\cite{protoDUNEPMTs}. However, the measured rate is higher than the expectation from these sources. A study~\cite{Luo:2020itx} suggests that Ar$_{2}^{+}$ drifting to the cathode could explain the high SPE rate recently measured in several experiments as these molecules recombine with electrons and neutralize with electronegative impurities producing photons. Unfortunately, this hypothesis cannot be tested in our detector because of the non-uniform drift field.

\subsection{Wavelength-shifting materials: PEN and TPB} 
\label{sec:wls}

The LAr scintillation light is produced at 127\,nm, a wavelength which most photosensors are not sensitive to, and fluorescent materials are introduced to shift the photon wavelength towards the visible range. ProtoDUNE-DP uses PMTs either covered with polyethylene naphthalate (PEN) foils or directly coated with tetraphenyl butadiene (TPB). While TPB is broadly used, PEN is a novel material, never used before in such a large scale experiment and whose efficiency is not well known. As TPB needs complex coating setups~\cite{Bonesini_2018}, the potential benefit of PEN comes from its simple handling, because PEN foils are flexible plastic sheets easy to fabricate and install. The PEN sample used in ProtoDUNE-DP is transparent and biaxially oriented, manufactured by GoodFellow \cite{GoodFellow_PEN}.
It has been installed as round disks of 240\,mm diameter and 0.125\,mm thickness placed over the top of the PMT glass surface, as shown in Fig.~\ref{fig:pds}. TPB was deposited over the PMT polished surface, using a dedicated evaporation system developed by the ICARUS experiment~\cite{Bonesini_2018}. The coating density is 0.2\,mg/cm$^2$, which corresponds to a coating thickness around 0.2\,µm. Both PEN and TPB have a maximum of re-emitted photons around a wavelength of 430\,nm~\cite{PEN-DMary,TPB-Francini2013}.


Photons produced after cosmic particles interact with the LAr will arrive at the wavelength-shifter (either PEN or TPB) and will convert into visible light that can be detected by the PMTs. The relative photon detection efficiency of the PEN-foil PMTs versus the TPB-coated PMTs, $NPE_\mathrm{PEN}/NPE_\mathrm{TPB}$, is experimentally determined by comparing the amount of light (in PEs) detected by a pair of PEN-TPB PMTs placed symmetrically with respect to the detector and the light source. An homogeneous response for the whole photocathode is considered.

Dedicated data-sets were taken selecting events with a signal amplitude larger than 13\,PEs in the trigger PMT, a TPB-coated PMT placed at the center of the detector (channel 16 in Fig.~\ref{fig:diagram}). The PMT pairs are selected among PMTs symmetrically placed near the trigger PMT (for example, channels 17 and 23). As the event position is not known, the number of photons reaching the PMTs cannot be compared on an event by event basis, but it can be assumed that the amount of photons reaching both PMTs is on average the same. The cosmic-muon flux is assumed to be isotropic.

The response of five PMT pairs is compared at different gains (10$^7$, 2$\cdot$10$^7$, 5$\cdot$10$^7$ and 10$^8$). The trigger configuration was kept constant in all data-sets, with the same gain (5$\cdot$10$^6$) and threshold (13\,PEs in amplitude) in the trigger PMT in order to study the same event sample at different gains.
To avoid saturation of the ADC and guarantee linear response of the PMTs, events with a signal amplitude larger than $\sim$30\,PEs on the trigger PMT are not included. This selection reduces the fraction of saturating events below 1\% in all PMTs in all data-sets.

The average light collected on the PMTs for the selected events is $\sim$200\,PEs on TPB-coated PMTs, and $\sim$50\,PEs on PEN-foil PMTs. The $NPE_\mathrm{PEN}/NPE_\mathrm{TPB}$ ratio is stable for each pair at different gains and different ranges. An average $NPE_\mathrm{PEN}/NPE_\mathrm{TPB}$ ratio of $0.25\pm0.03$ is obtained, as reported in Table~\ref{tab:WLSResults}. The error is the STD among PMT-pairs, which agrees with the expected error (0.03) due to the QE variation between 3 PMTs measured by the manufacturer. On average, ProtoDUNE-DP TPB-coated PMTs detect four times more photons than PEN-foil PMTs. This ratio is computed for a particular sample of muons; selecting events with a different track topology would vary this ratio.


\begin{table}[!ht]
\begin{center}
\caption{First row: Value of the relative photon detection efficiency, $NPE_\mathrm{PEN}/NPE_\mathrm{TPB}$. The error represents the STD for all the PMT pairs. Second row: relative WLS efficiency of PEN and TPB, $\epsilon_\mathrm{PEN}/\epsilon_\mathrm{TPB}$.}
\label{tab:WLSResults}
\begin{tabular}{cc}
\hline
  Parameter & Value  \\
    \hline
    $NPE_\mathrm{PEN}/NPE_\mathrm{TPB}$ &  $0.25\pm\,0.03$ \\
    $\epsilon_\mathrm{PEN}/\epsilon_\mathrm{TPB}$ & $0.35\pm0.09$ \\
    \hline
\end{tabular}

\end{center}
\end{table}

Additionally, a simple model is proposed to compute the relative WLS efficiency of the two materials. The number of detected photoelectrons is given by:
\begin{equation}
\centering
NPE =  \gamma \cdot \epsilon \cdot \Delta \cdot QE ,
\label{eq:PENTPB_NPE}
\end{equation}
where $\gamma$ is the number of VUV photons arriving to the wavelength shifter, $\epsilon$ is the WLS efficiency, $\Delta$ is the photon transport losses from the WLS to the PMT, and $QE$ is the PMT quantum efficiency. Then, the relative conversion efficiency can be derived:

\begin{equation}
\frac{\epsilon_\mathrm{PEN}}{\epsilon_\mathrm{TPB}} = \frac{NPE_\mathrm{PEN}}{NPE_\mathrm{TPB}} \cdot \frac{\gamma_\mathrm{coat}}{\gamma_\mathrm{foil}}\cdot \frac{\Delta_\mathrm{coat}}{\Delta_\mathrm{foil}}.
\end{equation}

The ratio of photons arriving to the TPB coating over the PEN foil is, on average, $\gamma_\mathrm{coat}/\gamma_\mathrm{foil} = 0.69\pm0.16$, as calculated using the cosmic-muon simulation described in section~\ref{sec:sim}. This factor depends on the selected sample, and the simulation of a different event topology would vary its value. The fact that the PEN foil receives 30\% more photons than the TPB coating can be explained as its two faces are exposed to LAr while the TPB only has one.

Considering an isotropic re-emission of the TPB coating, only 50\% of the photons will  reach the photocathode ($\Delta_\mathrm{coat}$ = 0.5). For the PEN foil, the transport losses are simulated considering an isotropic re-emission in the foil, and it is found that $\Delta_\mathrm{foil}$ = 0.247, meaning that only 25\% of the re-emitted photons will reach the photocathode due to the geometrical configuration of the foil with respect to the photocathode.

As a result, considering the geometrical differences and the measured ratio $NPE_\mathrm{PEN}/NPE_\mathrm{TPB}$, the relative WLS efficiency of both materials is estimated to be $\epsilon_\mathrm{PEN}/\epsilon_\mathrm{TPB}=0.35\pm0.09$, as shown in Table~\ref{tab:WLSResults}. TPB produces three times more visible photons than PEN, for the same amount of incident VUV photons. This agrees with the value of $0.34\pm0.01$ reported in \cite{PEN-Texas} for the same PEN sample. 

To introduce the PMT response in the simulation, the effective PMT photon-detection efficiencies at 127\,nm for TPB-coated PMTs ($E_\mathrm{TPB}$) and for PEN-foil PMTs ($E_\mathrm{PEN}$) are estimated. This effective efficiency provides the amount of photoelectrons detected per incident VUV photon, and it can be calculated by:

\begin{equation}
\centering
E= \epsilon \cdot \Delta \cdot \text{QE} \, . 
\end{equation}

The PMT quantum efficiency (QE) was measured by the manufacturer for three of the PMTs at room temperature. They measured a value of QE = $0.183\pm0.013$ at 430\,nm, and it is assumed that the PMT QE is stable when going to cryogenic temperature~\cite{Bueno_2008,Zhao_2021_PMT_QE_CT}. Since the values of TPB efficiency at 127\,nm reported in the literature show a large dispersion \cite{TPB-Benson2018,TPB_Graybill,Lally_TPB}, a 100\% TPB efficiency is assumed as in a previous work~\cite{311light} ($\epsilon_\mathrm{TPB}$=1). Then, assuming the PEN efficiency obtained by the relative performance of both systems in Table~\ref{tab:WLSResults} ($\epsilon_\mathrm{PEN} = 0.35\pm0.09$), the corresponding effective efficiency for TPB-coated and PEN-foil PMTs are $E_\mathrm{TPB}=0.09$ and $E_\mathrm{PEN}=0.016$, making the TPB coated-PMT six times more efficient. However, positioning the PEN foil directly over the PMT glass, as the TPB coating, would double its effective efficiency. These efficiencies are used in the analyses presented in sections~\ref{sec:light} and \ref{sec:muon}. 


\subsection{LAr scintillation time profile}
\label{sec:time}

The scintillation light emission in LAr has a characteristic time dependence as mentioned in section~\ref{sec:intro}. To get the scintillation decay times from the PMT waveforms, signals from cosmic muons are selected by triggering on a TPB-coated PMT with a minimum amplitude of 25\,PEs. The PMT gain is set at 5$\cdot$10$^{6}$ to minimize the ADC saturation, and events saturating the PMT~\cite{Lastoria:2748990} are excluded. Eleven of the PMTs suffer this phenomenon more frequently and are excluded for this analysis. 

\begin{figure}[ht]
    \centering
    \includegraphics[width=0.45\textwidth]{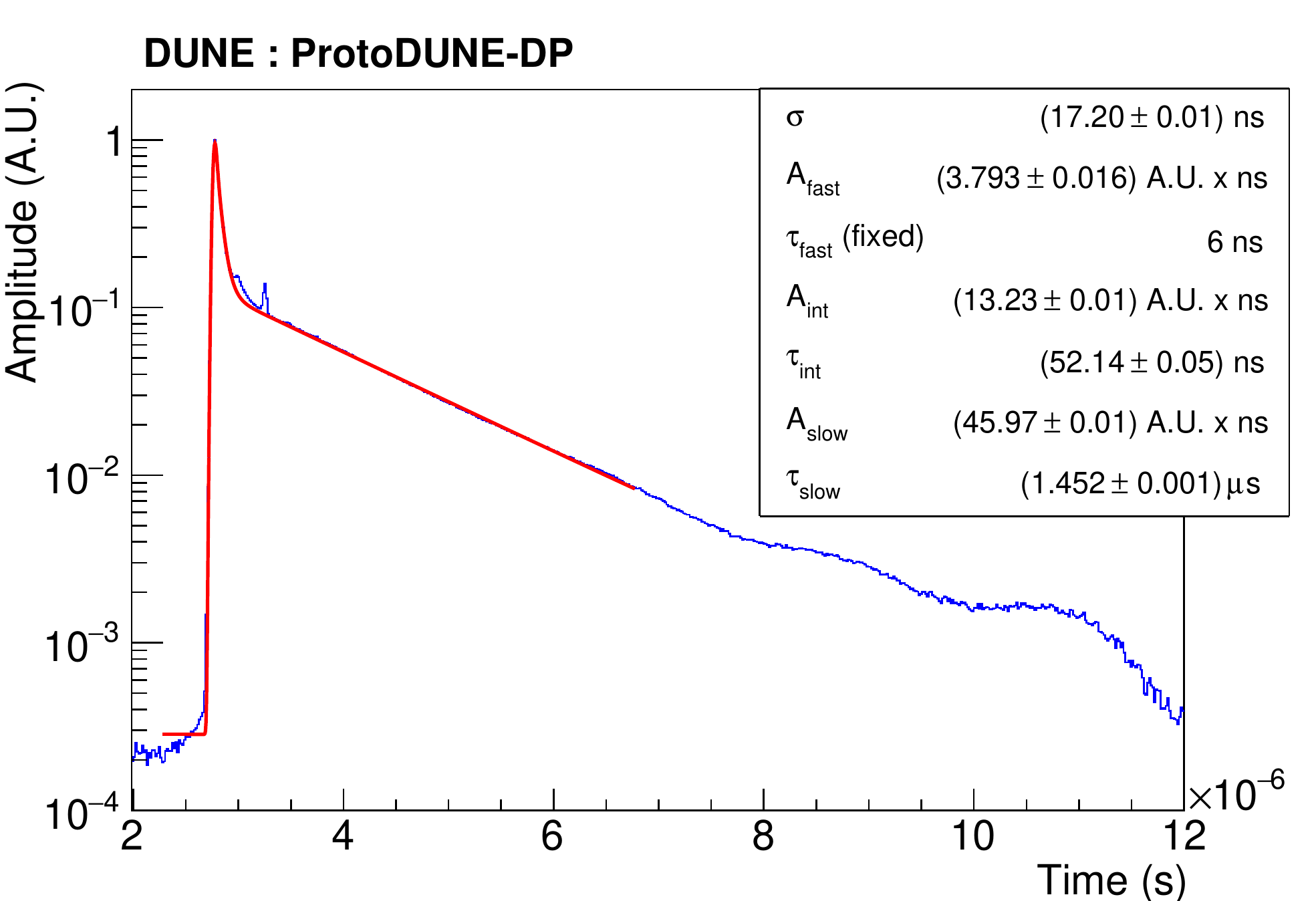}
    \caption{Example of the average scintillation waveform of a PEN PMT (blue). The fit to Eq.~\ref{eq:FitFunction} is shown in red. The corresponding parameters are shown in the legend.}
    \label{fig:Purity_ExampleFit}
\end{figure}

An average time profile is generated  for each PMT. The average waveform in the absence of drift field for one PMT is shown in Fig.~\ref{fig:Purity_ExampleFit}. Waveforms are well described by Eq.~\ref{eq:FitFunction}:

\begin{equation*}
f(t) = \sum_{j=\text{fast,slow,int}} \frac{2 A_{j}}{\tau_{j}} \exp\left[\frac{\sigma^{2}}{2 \tau_{j}^{2}} - \frac{t -t_{0}}{\tau_{j}}\right] \times
\end{equation*}
\begin{equation}
\times \left(1-\text{Erf}\left[\frac{\sigma^{2} - \tau_{j}(t-t_{0})}{\sqrt{2}\sigma \tau_{j}}\right]\right)
\label{eq:FitFunction}
\end{equation}

a sum of three exponential functions convoluted with a Gaussian function to represent the detector response. Although the scintillation time profile should in principle have only two components, from the decay to ground state of singlet ($\tau_\mathrm{fast}$) and triplet ($\tau_\mathrm{slow}$) argon excimers, an intermediate component ($\tau_\mathrm{int}$) is added in order to improve the fit as reported also by other LAr experiments~\cite{311light}. Given the 16-ns digitization sampling, the fit has a limited sensitivity to $\tau_\mathrm{fast}$ and this parameter is fixed to 6\,ns~\cite{Hitachi}. Additionally, two signal reflections at the flange feed-through appear $\sim$200\,ns and $\sim$400\,ns after the maximum, affecting the sensitivity to the $\tau_\mathrm{int}$ measurement. To mitigate this, the bins containing the reflections are excluded from the fit.

\begin{figure}[ht]
    \centering
    \includegraphics[width=0.45\textwidth]{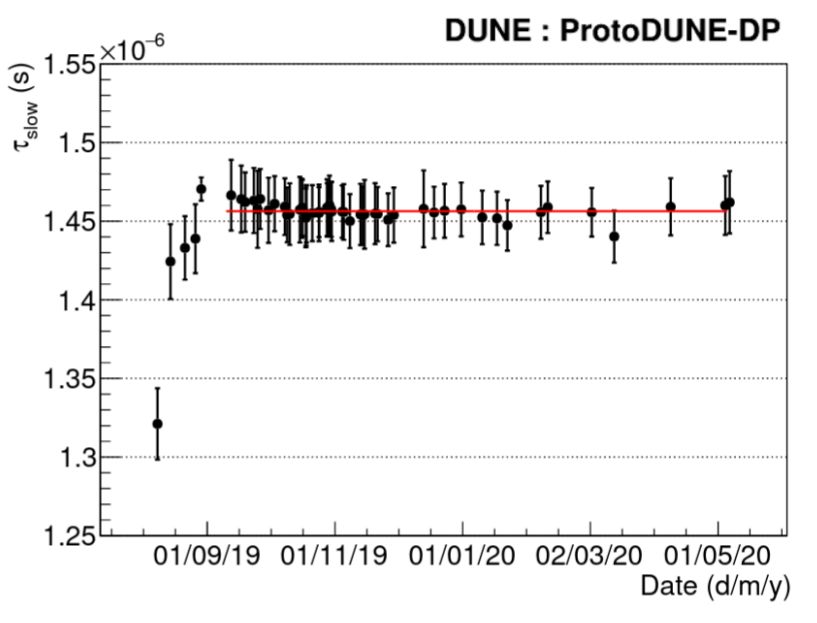}
    \caption{Monitoring of the $\tau_\mathrm{slow}$ component of the scintillation time profile during the detector operation. The error bars show the STD of the measurements for the different PMT waveforms. Red line at $\tau_\mathrm{slow}=1.46~\mu$s shows the average from September 2019 when stable operation began. }
    \label{fig:Purity_Monitoring}
\end{figure}

Purity is critical in LArTPCs, since impurities can reduce the signal by trapping the ionization electrons. The purity can be monitored using the PDS by measuring the lifetime of the triplet molecular argon excimers, $\tau_\mathrm{slow}$. Fig.~\ref{fig:Purity_Monitoring} shows the evolution of the average $\tau_\mathrm{slow}$ during the operation of ProtoDUNE-DP. The purity improved when the LAr purification system was turned on, and remained stable during the whole operation. The value of $\tau_\mathrm{slow}$ is $1.46 \pm 0.02$\,$\mu$s, with the error corresponding to the STD among the PMT waveforms. This average value has a small variation over time of just $0.004$\,$\mu$s. The absolute value indicates a high LAr purity at the ppb level. No significant difference is observed in $\tau_\mathrm{slow}$ between PEN and TPB PMTs. On average, $\tau_\mathrm{slow}$ is $1.45\pm0.02\,\mu$s on PEN PMTs and $1.46\pm0.02\,\mu$s on TPB PMTs.
    
\begin{figure}[ht]
    \centering
    \includegraphics[width=0.45\textwidth]{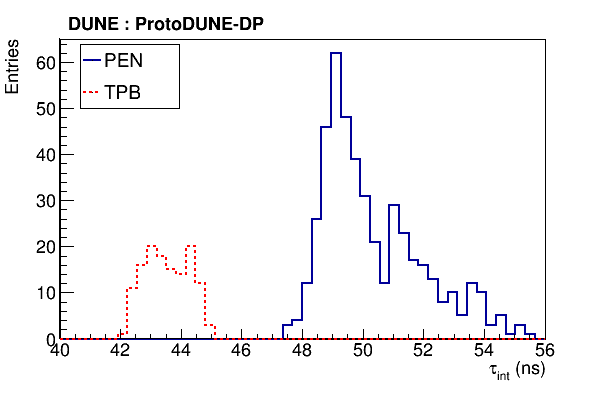}
    \caption{$\tau_\mathrm{int}$ distribution obtained from the fit to the average waveform for PEN (blue) and TPB PMTs (red).}
    \label{fig:PENTPB_TauInt}
\end{figure}

Fig.~\ref{fig:PENTPB_TauInt} shows the value of $\tau_\mathrm{int}$ for PEN and TPB PMTs obtained from the fit of the average waveforms using data sets for which the purity was already stable, from September 2019 to May 2020. An average value of $50.3 \pm 1.7$\,ns is obtained for the PEN PMTs, and a faster response of $43.6 \pm 0.7$\,ns for the TPB PMTs. The clear difference between the two different WLS points to a delayed emission time by the WLS material, as proposed in~\cite{Segreto, Whittington}. 

\section{Cosmic-muon light simulation} 
\label{sec:sim}

ProtoDUNE-DP light simulations are carried out using LArSoft~\cite{LArSoft}, a physics software package designed for LAr neutrino experiments, such as DUNE. 

The detailed geometry of the ProtoDUNE-DP detector has been implemented in Geant4~\cite{geant4} using geometry description markup language (GDML) files. It includes the full cryostat with its support structure,  cathode, field cage, LEMs, PMTs and ground grid.

The simulation of the cosmic-induced particles reaching ProtoDUNE-DP is based on CORSIKA (COsmic Ray SImulation for KAscade)~\cite{CORSIKA}, a Monte-Carlo (MC) simulation package. It is based on a multi-component model of primary cosmic rays, the constant mass composition (CMC) model~\cite{forti}, for a more complete modeling of the flux. CORSIKA generates showers from each specific cosmic ray type (Fe, He$_2$, Mg, proton) according to a power law distribution of the primary particle energy. The FLUKA model~\cite{FLUKA} describes the hadronic shower propagation. The LArSoft simulation draws from a random set of pre-generated air showers from a specific database created with CORSIKA. The number of showers per event and per primary particle source is extracted according to a Poisson distribution around the predicted average number of primary cosmic rays for each source. The detector location (CERN), altitude and latitude, is used. The start time and spatial origin of the cascades of secondary particles are chosen randomly, and time and space correlations of the particles within each shower are preserved.

Muons crossing the CRT panels of ProtoDUNE-DP are also simulated. Single muons are generated with a defined initial position, momentum, and momentum spread. The energy distribution of the muons is taken from CORSIKA for a more realistic outcome. The data-driven track entry/exit distributions in the CRT panels are taken as inputs to define the track topology of the events. A topological selection is carried out and only muons with a complete trajectory between the CRTs are accepted, as deviations from the ideal straight trajectory can appear during the propagation through the detector and muons can decay along the track length.

The cosmic muons simulated with the event generators enter the LAr volume and deposit energy along their tracks as they interact with LAr. An incident muon behaves as a minimum ionizing particle (MIP) and deposits about 2 MeV/cm. First, the simulation of the muon energy deposition is performed in Geant4. Then, the number of scintillation photons is computed by multiplying the deposited energy by a light yield of 4$\cdot 10^4$ scintillation photons per MeV in the absence of a drift field.  The time profile is simulated with an exponential fast decay of 6\,ns for 30\% of the photons, and an exponential slow decay of 1590\,ns for the other 70\% of the photons. Simulations were carried out before obtaining the ProtoDUNE-DP results described in section~\ref{sec:per} and no intermediate component is included. No drift field is assumed in this simulation to match the data taking conditions. Note also that only cosmic particles are simulated (only muons in the CRT case), and backgrounds from natural radioactivity are not currently included in the simulation. 

The photon propagation in LAr, from the production point to the PMT array, is performed with Geant4 in LArSoft. The VUV-light attenuation due to absorption by impurities in LAr is simulated (20-m absorption length, equivalent to 3\,ppmv of nitrogen contamination in LAr~\cite{Jones:2013bca}) as well as the Rayleigh scattering length (RSL) (99.9\,cm as baseline value~\cite{Babicz:2020den}, but also a shorter length of 61.0\,cm~\cite{GRACE2017204} is tested). A VUV reflectance of 26\% in aluminum is taken (field cage)~\cite{ICARUStechnote}, and the same VUV reflectance is assumed for stainless steel (SS) surfaces (cryostat walls, cathode and ground grid)~\cite{Zatschler:2020yjp}. Full absorption is considered for the rest of the materials. Table~\ref{tab:sim} summarizes the relevant parameters of the light propagation used in the simulation.

\begin{table}[!ht]
\begin{center}
\caption{Parameters of the light propagation simulation in ProtoDUNE-DP.}
\label{tab:sim}
\begin{tabular}{ccc}
\hline 
Parameter & Baseline value & Alternative \\
\hline 
Absorption length & 20\,m & - \\
\hline 
RSL &  99.9\,cm & 61.0\,cm \\
\hline 
VUV Reflectance & 26\% in Al \& SS, & 0\% for all \\
 & 0\% for the rest &   \\
 \hline 
Voxel size & 0.34\,$\times$\,0.32\,$\times$\,0.34\,m$^3$ & -\\
\hline 
\end{tabular} 
\end{center}
\end{table}

\begin{figure*}[ht]
    \centering
    \includegraphics[width=0.88\textwidth]{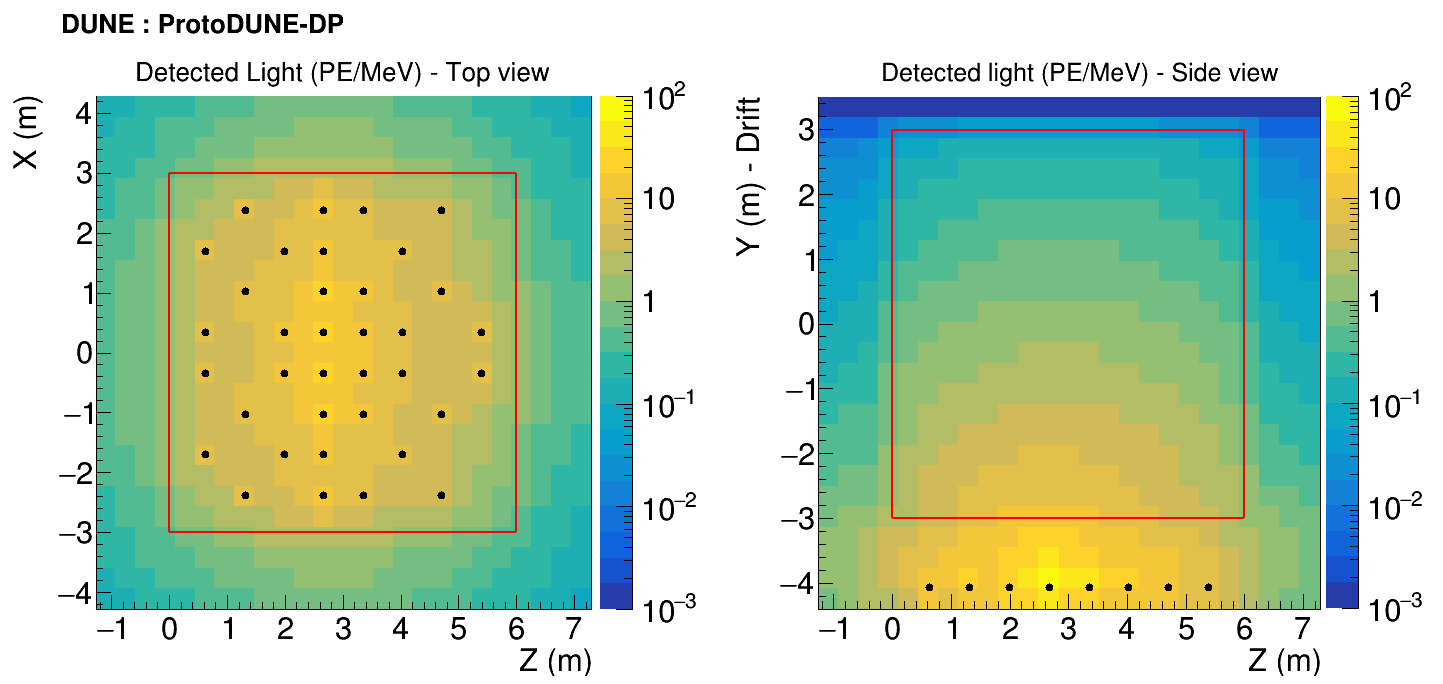}
    \caption{Maps of the detected light (sum of PMT visibilities multiplied by efficiency, $E_\mathrm{TPB}=0.09$ and $E_\mathrm{PEN}=0.016$) per voxel for the baseline photon library in ProtoDUNE-DP: 4$\cdot 10^4$\,photons per MeV, 99.9\,cm of RSL and 26\% VUV reflections. The PMT positions are represented with black dots and the active volume with a red rectangle.}
    \label{fig:lightmaps}
\end{figure*}

The complete simulation of the propagation of all the photons produced by each particle crossing the detector would require an enormous amount of CPU time. Hence, pre-generated libraries are employed to simulate the light propagation in an efficient way: a Geant4-driven simulation is generated and the results are stored and parametrized in the so-called photon libraries. The LAr volume is divided into 3D voxels and 10$^8$ photons per voxel are isotropically and uniformly generated. A photon library contains, for each PMT-voxel combination: the visibility (photon detection probability), the minimum time for the light to reach the PMT (arrival time), and the Landau-fit parameters of the propagation time distribution. In order to mitigate the relatively large size of the voxels, 3D interpolations among them are used at each step of the track.

Three different photon libraries are generated to study the impact of several parameters in ProtoDUNE-DP: (A) the baseline photon library with 99.9-cm RSL and 26\% VUV-light surface reflectance, (B) a library with 61-cm RSL and 26\% of surface reflectance, and (C) a library with 99.9-cm RSL and no reflection. The maps with the detected light per deposited MeV across the volume for the baseline photon library (A) are displayed in Fig.~\ref{fig:lightmaps}.

The PMT response is simulated with a dedicated module that produces a waveform for each PMT. Each 127-nm photon arriving to the WLS has a probability to produce a photoelectron in the PMT photocathode. The WLS-PMT photon-detection efficiency includes the WLS efficiency, the propagation of the visible light from the WLS towards the photocathode, and the PMT quantum efficiency. The values used in the light simulation are taken from studies presented in section~\ref{sec:wls}: $E_\mathrm{TPB}=0.09$ and $E_\mathrm{PEN}=0.016$.

Waveforms are produced by adding the SPE response for each detected photon. The simulated PMT response is linear, and includes a dark current component of 1.7\,kHz~\cite{Belver:2019lqm}). The waveform digitization considers a 16-ns sampling matching the data acquisition system.


\section{Light production and propagation in LAr measured with cosmic muon data}
\label{sec:light}

The pure LAr of ProtoDUNE-DP is an optimal medium to study the scintillation light production, propagation and collection in a LAr volume. The experiment provides valuable technical feedback for future light detection systems as well as results on the scintillation light mechanisms in LAr. In this section, PMT data acquired with the CRT-trigger system are analyzed to profit from the off-line reconstruction of the track trajectory that such a trigger allows. Muon tracks crossing the detector diagonally and in the downward direction are selected. Information about the distance the light travels from its production point in the LAr volume to the PMT detecting the signal is retrieved.

\begin{figure}[ht]
    \centering
    \includegraphics[width=0.35\textwidth]{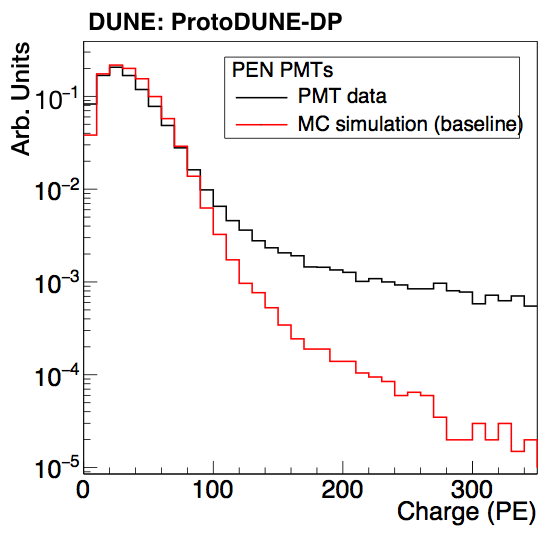}
    \caption{Normalized distributions of the collected S1 charge for CRT triggered light data and simulation.}
    \label{fig:Q1D}
\end{figure}

The analysis focuses on the 30 PEN PMTs because they allow exploring a wider track-PMT distance range than the TPB PMTs. Consistent results are obtained with the TPB PMTs. The S1 charge (number of PEs) of each triggered signal is obtained by integrating the PMT waveform in a 4-$\mu$s window. An event selection is made in both data and simulation to obtain comparable event samples:

\begin{enumerate}
    \item Time of flight (data) / geometrical cut (MC): a data event is considered a good muon-candidate if the time of flight between the CRT panels (from top to bottom) is between 40 and 45\,ns, and only one scintillating bar per panel is triggered. This cut contributes to the rejection of background events like electrons, particle showers, muon bundles and fake triggers. In the simulation case, only CRT-trigger-like muons are kept by requiring that the muon-track trajectory crosses both CRTs. 
    \item Stable baseline before trigger: this cut intends to avoid the pile-up of signals. The waveform RMS is obtained in a 1-$\mu$s window before the signal trigger and the baseline is considered to be stable if the result is below 1\,ADC count.
    \item No ADC saturation: data events with waveforms saturating the ADC are excluded, both in data and MC.
    \item Maximum charge: the charge per event must be below 100\,PE for PEN PMTs. This limit is applied to suppress high-energy events (vertical showers instead of diagonal muons) that affect the distribution of interest in data but not in the MC as they are not simulated, see Fig.~\ref{fig:Q1D}.
\end{enumerate}

\begin{figure}[ht]
    \centering
    \includegraphics[width=0.4\textwidth]{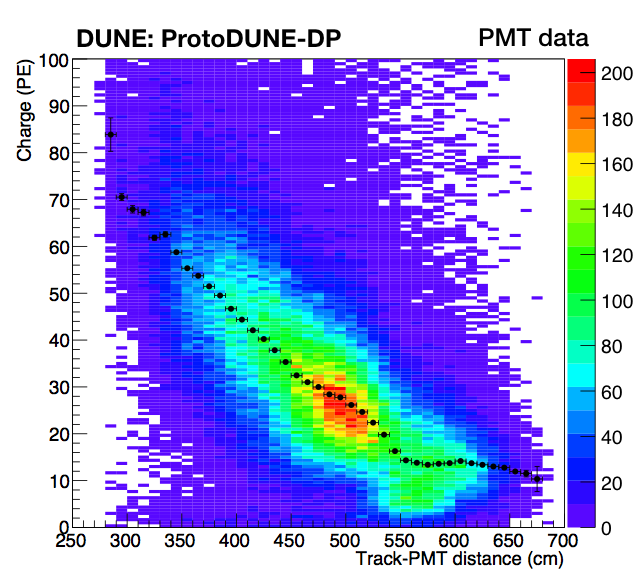}\\
    \includegraphics[width=0.4\textwidth]{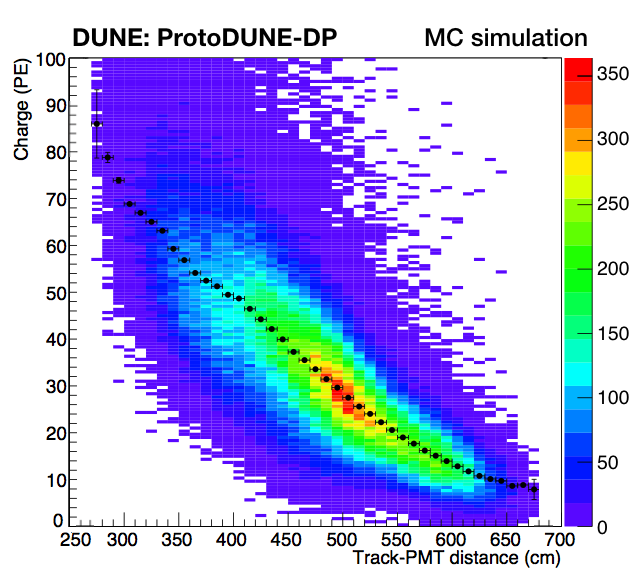}
    \caption{Collected S1 charge versus track-PMT distance for PEN PMTs: Light data (top) and MC sample (bottom). The color maps contain all the S1 signals detected by the PMTs and passing the event selection. A Gaussian fit of each charge-distribution every 10\,cm is performed and the mean values are plotted in black over the map. The vertical error bars correspond to errors from the fits and a 5-cm distance uncertainty (horizontal error bars) is included.}
    \label{fig:Qvsdistance}
\end{figure}

Since the propagation effects on the collected charge by the PMTs depend on their distance to the muon track, analyses shown in this section are mainly based on the correlation between the S1 signal (S1 charge in PE units) and the minimum distance from the muon track to the detecting PMT (simply referred to as track-PMT distance). Fig.~\ref{fig:Qvsdistance} shows the 2D distribution of theses variables for data and MC. A good agreement is observed up to $\sim$5\,m; beyond that distance, the correlation between charge and distance is lost in data, presumably, because low-energy background (only present in data and contributing with up to 8-9\,PE per integration window) prevails over the signal. The profile histogram superimposed on the 2D plot corresponds to a Gaussian fit of the charge-distribution every 10\,cm. In the case of the data samples, an additional systematic error of 4\% (determined by varying the PMT gain) is added in quadrature to the fit error of the charge.

\begin{figure}[ht]
    \centering
    \includegraphics[width=0.4\textwidth]{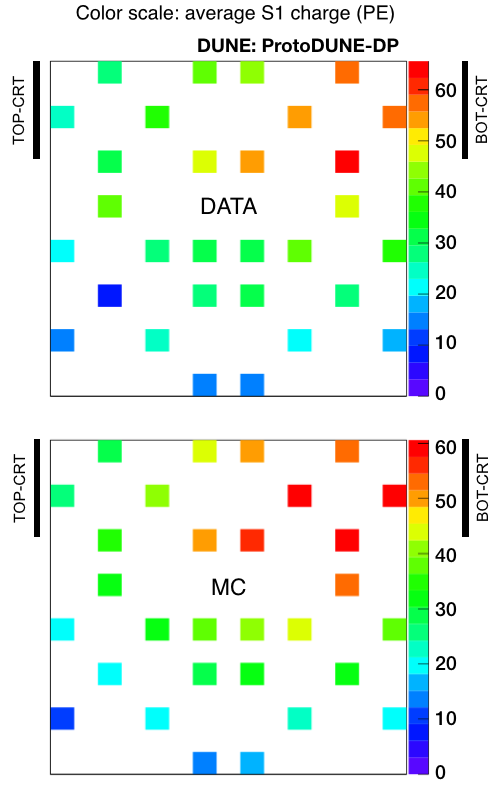}
    \caption{Average S1 charge per channel (in PE) in CRT-trigger mode: top view of PEN-PMT array for data (top panel) and simulation (bottom panel). Each color square represents one PMT channel at its position in the detector. The approximate positions of the CRT frames are indicated with black lines.}
    \label{fig:S1CRT}
\end{figure}

The top-view light detection maps (average S1 charge per optical channel) for the PEN PMTs both for data and MC are presented in Fig.~\ref{fig:S1CRT}. The obtained light detection patterns reproduce the expected gradient: the PMTs which are closer to the bottom CRT (top right corner in the top view) detect more light than the rest.

\subsection{Drift field effect on light production}

\begin{figure}[ht]
    \centering
    \includegraphics[width=0.4\textwidth]{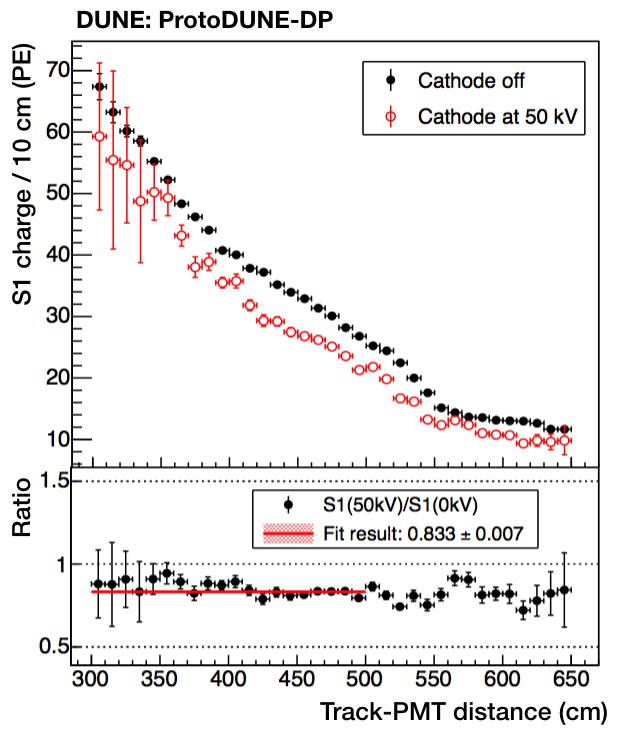}
    \caption{Top panel: Average S1 charge collected by the PEN PMTs as a function of the track-PMT distance without drift field (black) and with cathode at -50\,kV (red). The larger error bars at short distance are due to a lack of statistics. Bottom panel: Ratio between the two distributions. The ratio is fitted to a constant value (result: 0.833~$\pm$~0.007).}
    \label{fig:Fields}
\end{figure}

The reduction of the detected light with the increasing drift field is investigated. This reduction is due to the suppression of the electron-ion recombination by drift field, which reduces the primary scintillation light production.
The electric field in ProtoDUNE-DP is not uniform across the active volume, which makes infeasible a complete understanding of the drift field effect over the light production. Nonetheless, the light levels detected with the PDS without drift field and at the maximum operating cathode HV (-50\,kV) are compared to roughly quantify the light yield decrease. Fig.~\ref{fig:Fields} shows the charge-distance distributions at the two cathode HVs and the corresponding ratios between them. The ratio as a function of the track-PMT distance is fitted to a constant to obtain the average ratio of 0.833\,$\pm$\,0.007, which means that at least 17\% of the scintillation light detected in the absence of a drift field comes from electron-ion recombination.

\begin{figure}[ht]
    \centering
    \includegraphics[width=0.4\textwidth]{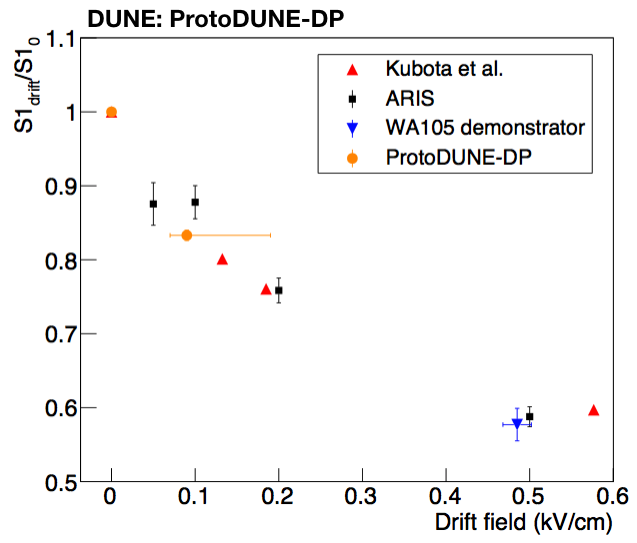}
    \caption{Light yield reduction with drift field measured by various experiments (red triangles from~\cite{Kubota}, black squares from~\cite{Aris}, blue inverted triangles from~\cite{311light}), and the ProtoDUNE-DP result (orange circles) discussed in this section.}
    \label{fig:DriftField}
\end{figure}

The CRT-trigger muons cross different fields, which leads to a difficult quantification of the overall drift field responsible for the light level reduction. An estimate of the effective electric field along the CRT-trigger muon tracks was made using the 3D simulation of the drift field in ProtoDUNE-DP corresponding to a cathode HV of -50\,kV, see Fig.~\ref{fig:mapfield}. The average field value obtained for these tracks is 0.09$^{+0.10}_{-0.02}$\,kV/cm, where the errors are determined asymmetrically, as the RMS of the values above and below the mean value separately. The reduction of the S1 signal for this drift field is plotted in Fig.~\ref{fig:DriftField} and follows the empirical Birks’ law. Despite the relatively large uncertainty of the ProtoDUNE-DP result, a fair agreement is found with the literature for both ground-level cosmic muons \cite{311light} and MeV-electrons \cite{Kubota,Aris}.

The scintillation time profile is also dependent on the drift field. The two excited molecular states, singlet and triplet, are formed either through recombination or excitation with different probability, in such a way that the normalization constants, $A_\mathrm{fast}$ and $A_\mathrm{slow}$ of Eq.~\ref{eq:FitFunction}, are expected to change as a function of the drift field as the light emitted by electron-ion recombination is suppressed. 
According to~\cite{Kubota} the ratio $(A_\mathrm{fast}+A_\mathrm{int})/A_\mathrm{slow}$ is  expected  to  decrease but the recent measurement presented in~\cite{311light} shows an increase of 34\% at 500\,V/cm. ProtoDUNE-DP has observed an increase of 28\% at -50\,kV cathode voltage, corresponding to an average field value of 0.09$^{+0.10}_{-0.02}$\,kV/cm, which is consistent with the 23\% measured in~\cite{311light} at a similar electric field strength.

\begin{figure}[ht]
    \centering
    \includegraphics[width=0.45\textwidth]{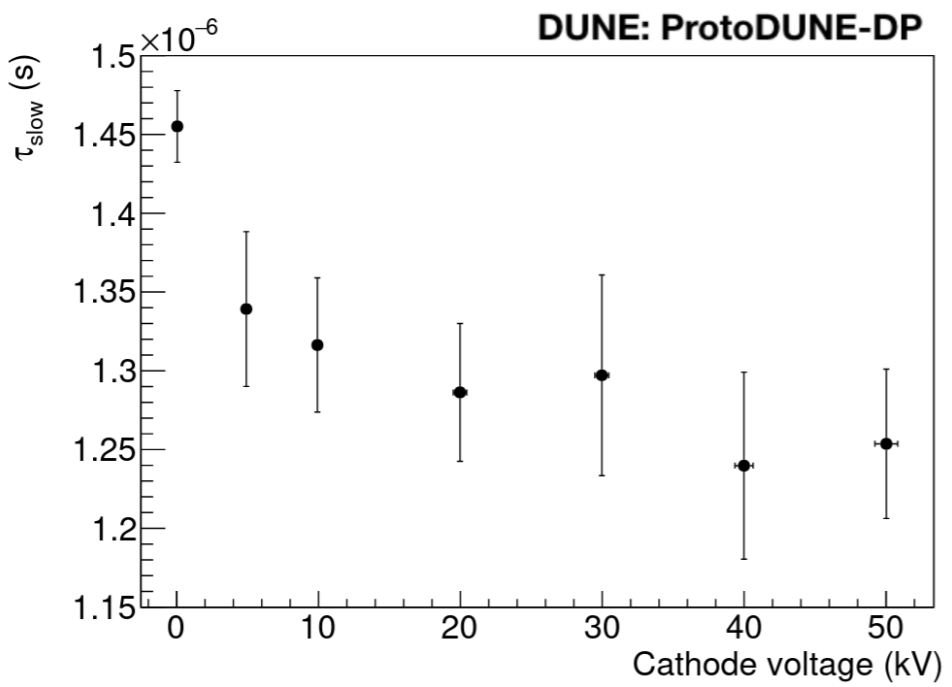}
    \caption{Evolution of $\tau_\mathrm{slow}$ with the cathode voltage for CRT-trigger events.  The results are obtained averaging PEN PMTs and the error bars indicate the STD among them.}
    \label{fig:slowvsE}
\end{figure}

A decrease of $\tau_\mathrm{slow}$ with the drift field was reported for the first time in~\cite{311light} and is also observed in ProtoDUNE-DP, see Fig.~\ref{fig:slowvsE}. A model is proposed in~\cite{Segreto:2020qks}, taking into account the quenching of the long-lived triplet states through the self-interaction with other triplet states or through the interaction with molecular Ar$^+_2$ ions. It successfully explains the experimentally observed dependence of $\tau_\mathrm{slow}$ with the intensity of the applied electric field.

\subsection{Light propagation}
\label{sec:lighprop}

The size of ProtoDUNE-DP, the longest drift-distance LArTPC ever operated, allows for an unprecedented study of the light propagation. The Rayleigh scattering length (RSL) can impact on the amount of light collected. An evaluation of the RSL value is carried out by comparing the measured light signals with the light predicted by the MC simulation testing two lengths (61.0\,cm~\cite{GRACE2017204} and 99.9\,cm~\cite{Babicz:2020den}) obtained in experimental measurements. Modeling the dependence of the light attenuation from the track-PMT distance with a decaying exponential function allows the measurement of the overall effective attenuation in data and MC and the evaluation of the agreement for the different simulated configurations. The track-PMT distance range studied is 4-5\,m to focus on the longer distances where the sensitivity to the RSL effect is higher.

\begin{figure}[ht]
    \centering
    \includegraphics[width=0.4\textwidth]{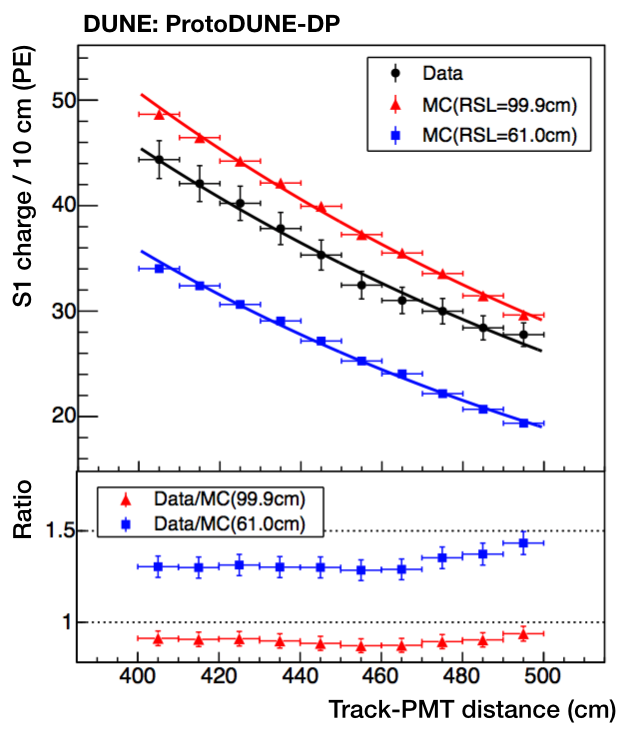}
    \caption{Top panel: Average S1 charge collected by the PEN PMTs as a function of the track-PMT distance. Two MC samples with different RSL values (61.0\,cm and 99.9\,cm) are compared to data. For each distribution, an exponential fit is 
    plotted as a solid line. Bottom panel: Data-MC ratio for the two previous MC samples.}
    \label{fig:Raileigh}
\end{figure}

In Fig.~\ref{fig:Raileigh}, each S1 charge-distance correlation under study is fitted to an exponential and the data-MC ratios for the two simulations are also presented. Looking at the distribution shape, the agreement between data and the 99.9-cm MC sample is better than with the 61.0-cm value. The attenuation length values obtained are presented in Table~\ref{tab:rsl}, and it is observed that the data value also agrees better with the 99.9-cm MC value. The attenuation length accounts for the effective attenuation of the light in ProtoDUNE-DP, but it is not a physical property of the LAr as it depends, among other factors, on the topology of the selected tracks and detector geometry. The measured attenuation length is higher than the RSL, so the light is expected to undergo Rayleigh scattering before being heavily attenuated due to, for example, absorption by LAr impurities or detector elements. This long light path before absorption is achieved in ProtoDUNE-DP thanks to the excellent LAr purity and the large free LAr volume with no nearby components.

\begin{table}[!ht]
\begin{center}
\caption{Attenuation length values obtained from the exponential fits shown in Fig.~\ref{fig:Raileigh}.} 
\label{tab:rsl}
\begin{tabular}{ccc}
\hline
Sample & $\lambda_\mathrm{att}$ (cm)\\
    \hline
Data   &  180\,$\pm$\,17\\
MC (RSL 99.9\,cm)  &  180\,$\pm$\,10\\
MC (RSL 61.0\,cm)  &  157\,$\pm$\,8\\
    \hline
\end{tabular}
\end{center}
\end{table}


By comparing MC samples generated with the photon libraries (A) and (C), the impact of the VUV-photon reflections on the light reaching the PMT array can be evaluated. It is found that 11\% of the light detected by PEN PMTs in the baseline MC corresponds to VUV light reflected on the field cage, cryostat walls, cathode and ground grid.

Finally, the light detection is studied through the effective TPB and PEN PMT efficiencies described in section~\ref{sec:wls}. The data-MC charge ratios obtained are 0.898\,$\pm$\,0.002 and 1.150\,$\pm$\,0.005 for PEN and TPB PMTs, respectively, in the selected track-PMT distance range. The result is that the simulation efficiency is validated within 10-15\% of data. The simulation underestimates the light detected by the TPB PMTs, which may be because not all the reflected light is simulated, or the TPB WLS efficiency may be under-estimated. However, the light detected by PEN PMTs in data is lower than in the simulations, which can be attributed to a PEN WLS efficiency over-estimation.
\section{Measurement of the cosmic muon rate and light yield}
\label{sec:muon}

The rate of cosmic particles crossing the detector is high because ProtoDUNE-DP is located on the surface. The high rate represents a perfect test bench to study the PDS capability for detecting muons and providing calorimetric information. 

The S1 rate and charge from cosmic muons crossing ProtoDUNE-DP as well as the muon flux are evaluated in section~\ref{sec:S1rate}. In addition, the observed light yield from muons is investigated for two different WLS methods in section~\ref{sec:PE/MeV}. 

In total, five data sets acquired over seven months with the random trigger are analyzed and compared with CORSIKA-based simulations. The cosmic-muon data sample allows the validation of the PDS simulation to demonstrate that the detector response is correctly modeled and understood. CRT-trigger data are also analyzed to evaluate the detected light yield for different muon samples.

\subsection{Measurement of the cosmic muon flux}
\label{sec:S1rate}

The results shown in this section are based on the analysis of S1 signals from light data acquired in random-trigger mode. These data comprise varied signals as the light reaching the PMTs can be generated at any location within LAr volume and be associated with very different track topologies and energies.

The muon S1 signals are identified in the PMT waveforms using a custom peak-finding algorithm which takes the following aspects into account:
\begin{enumerate}
    \item Any pulse in the waveform with an amplitude larger than the expected SPE amplitude plus 2$\sigma$ is considered a candidate S1 signal. The amplitude threshold is based on the results of the average SPE amplitude presented in section~\ref{sec:spe}. The lack of $\tau_\mathrm{int}$ in the simulated scintillation time profile (see section~\ref{sec:sim}) results in a different charge-amplitude correlation than in data and, hence, a correction of the amplitude is applied in the MC analysis in order to select events that correspond to the same deposited energy as      in data.
    \item After a candidate S1 signal, a veto time window of 4.8\,$\mu$s is applied. If another candidate S1 signal is detected during that time, a new 4.8-$\mu$s veto window beginning at that point is imposed.
    \item An optical reconstruction to reject small and uncorrelated signals is performed by requiring coincidences among PMTs. A coincidence occurs when at least two PMTs detect the candidate S1 signal within a time interval of 112\,ns. This cut is effective in rejecting candidate S1 signals at low energy in data, where the background, which is not considered in the MC, is dominant. About 40\% and 10\% of candidate S1 signals are rejected in data and MC, respectively.
\end{enumerate}

The rate of S1 signals detected by each PMT is then computed. It should be noted that the obtained S1 rate is not the overall cosmic muon rate in the detector but the average muon rate per PMT. Results presented in Table~\ref{tab:S1rate} show a higher rate of simulated S1 signals induced by muons: around 16\% (11\%) higher S1 rates are detected by the TPB (PEN) PMTs in simulation than in data. Nevertheless, the muon flux with the model chosen within CORSIKA to generate the primary cosmic particles, the CMC model, is expected to be up to 20\% above the one obtained with a model assuming only cosmic protons, as reported in~\cite{microboone2021}, and the flux can be up to 25\% lower with other particle generators. Thus, the observed data-MC deviation is within the discrepancy between the cosmic-ray generator models.



\begin{table}[!ht]
\begin{center}
\caption{Average S1 rate per PMT in random-trigger mode and data/MC ratio. The error of the S1 rate corresponds to the STD among PMTs.}
\label{tab:S1rate}
\begin{tabular}{cccc}
\hline
PMT & \multicolumn{2}{c}{S1 rate per PMT (kHz)} & Data/MC ratio\\
\hline
& Data & MC & \\
    \hline
TPB   & 8.8\,$\pm$\,0.5 & 10.2\,$\pm$\,0.5 & 0.86\,$\pm$\,0.09\\
PEN   & 5.5\,$\pm$\,0.6 & 6.1\,$\pm$\,0.4 & 0.90\,$\pm$\,0.10\\
    \hline
\end{tabular}
\end{center}

\end{table} 

\begin{figure}[ht]
    \centering
    \includegraphics[width=0.45\textwidth]{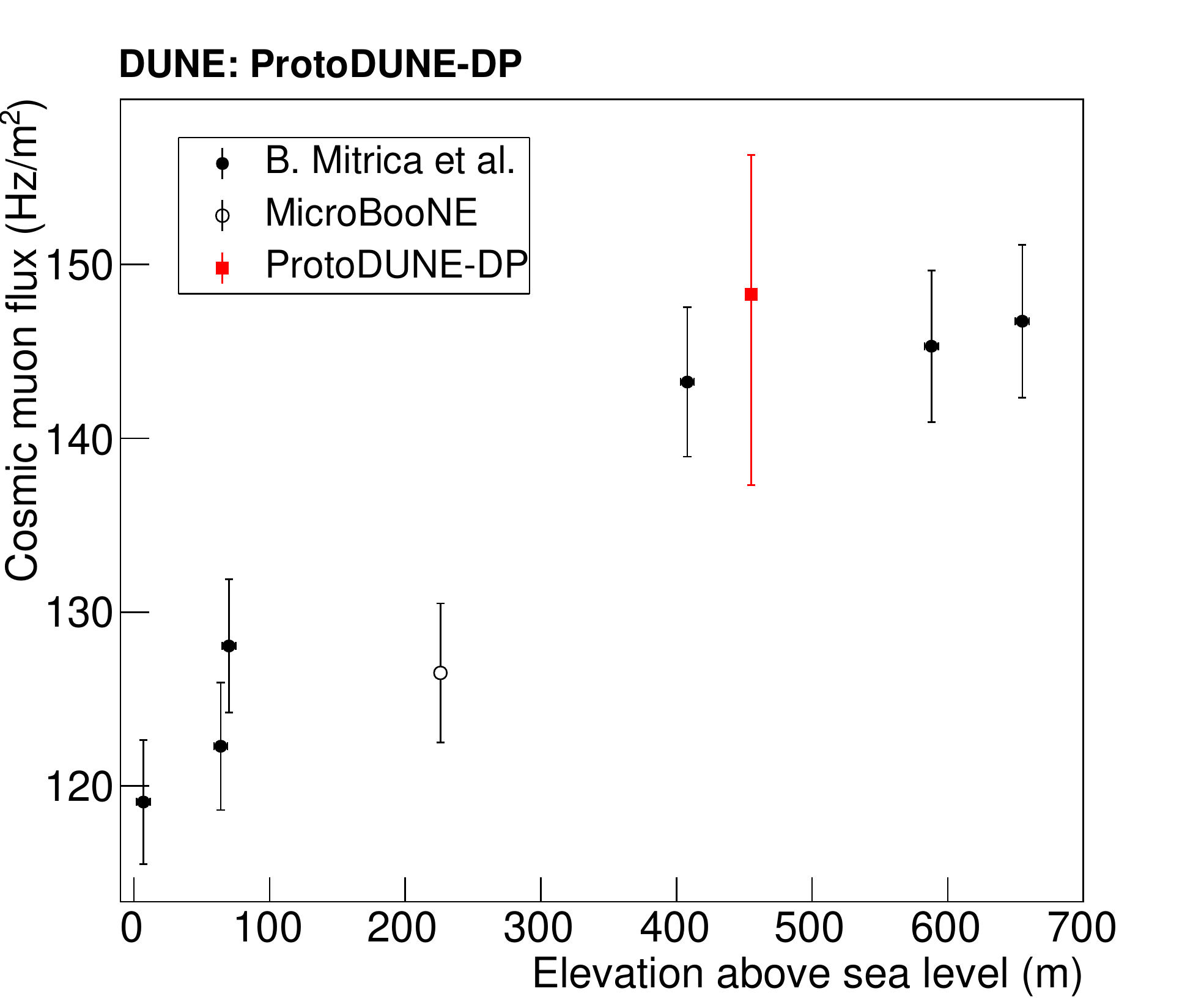}
    \caption{Cosmic muon flux determined in ProtoDUNE-DP at CERN at 455\,m above mean sea level (red square) compared to the measurements reported in \cite{microboone2021} (filled black dots) and \cite{mitrica2011} (open black dot).}
    \label{fig:flux}
\end{figure}

Finally, the atmospheric muon flux is assessed. Considering the surface covered by the PMT array and the fraction of muons crossing such an effective area in the MC, the predicted atmospheric muon flux at the Earth's surface by CORSIKA is 166\,Hz/m$^2$. Then, the predicted flux is scaled by the ratio between the data and MC rates given in Table~\ref{tab:S1rate}. A cosmic muon flux of 148$^{+8}_{-11}$\,Hz/m$^2$ is obtained in ProtoDUNE-DP at CERN (at 455\,m altitude above mean sea level). The systematic uncertainty on the flux is computed by varying the threshold in amplitude for the S1 identification by $\pm$20\% ($\sim$10-25~ADC) given the uncertainty in the MC waveform simulation, and the PMT efficiency in the MC by $\pm$10\% as concluded in section~\ref{sec:light}. The ProtoDUNE-DP cosmic muon flux result is found to be consistent with other measurements at ground-level given in the literature, as can be seen in Fig.~\ref{fig:flux}.


\subsection{Observed light yield from cosmic muons}
\label{sec:PE/MeV}


The light yield for muon interactions observed by the PDS is investigated considering the use of the system as a calorimeter. The event selection of random-trigger data explained in section~\ref{sec:S1rate} is applied. Good stability among data taken throughout seven months in terms of S1 rate and charge collection (2-5\% of STD among different sets in both cases) is observed, which indicates stable performance of the system.

\begin{figure}[ht]
    \centering
    \includegraphics[width=0.45\textwidth]{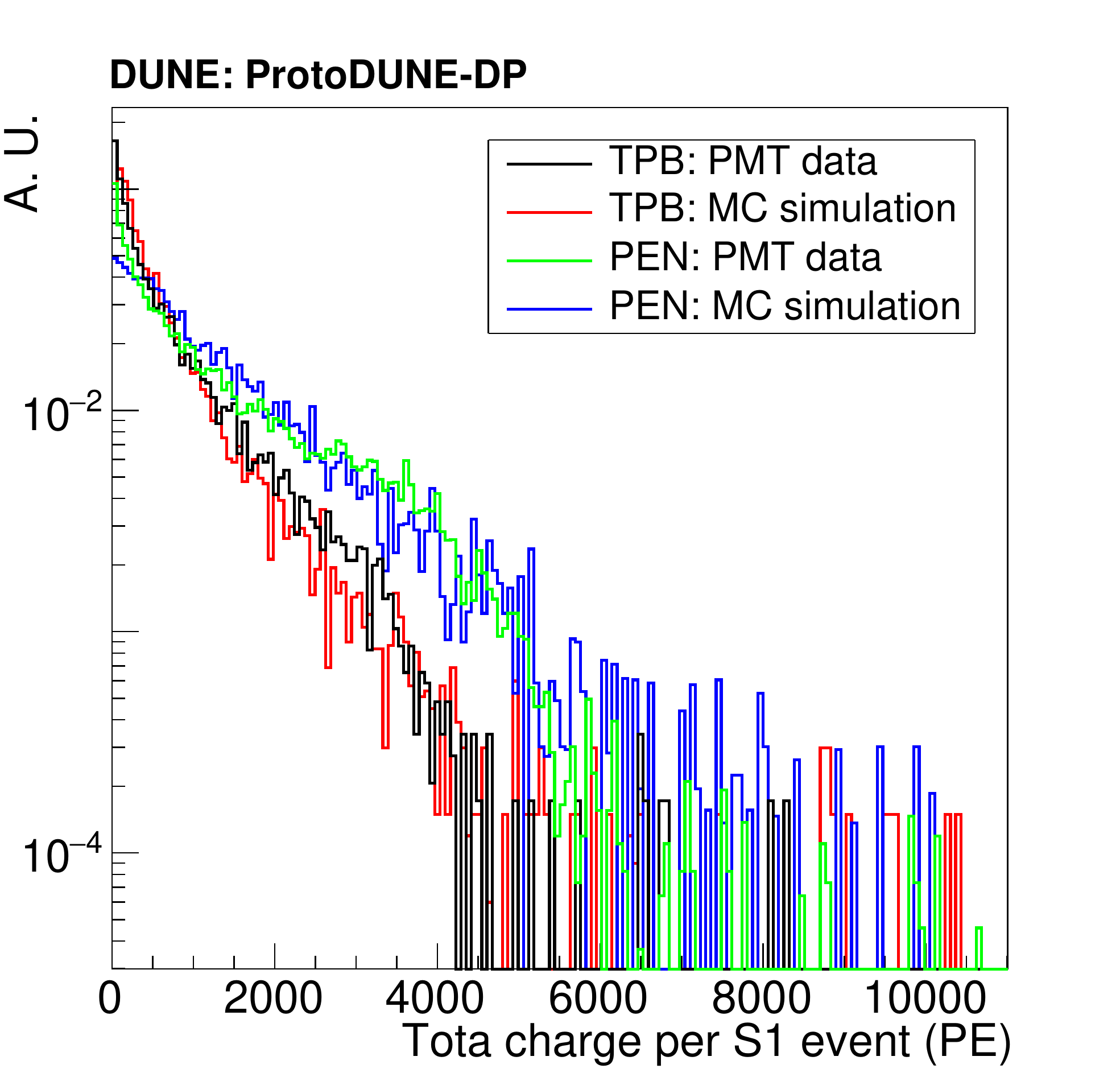}
    \caption{Normalized distributions of the total S1 charge per cosmic muon event (sum of all detecting PMTs) in random trigger. Both data and MC (TPB and PEN) cases are displayed. The ADC saturation in data that limits the charge range is fairly reproduced in the MC.}
    \label{fig:S1Qtotal}
\end{figure}

Taking into account all the cosmic muon tracks across the LAr volume, the average deposited energy per muon is determined with the CORSIKA cosmic-ray simulation to be 813\,MeV. Then, the observed light yield in random trigger mode is calculated as the total S1 charge collected by all PMTs normalized by the average deposited energy per muon. The results can be seen in Table~\ref{tab:PEMeV}. This approach is validated by the fair agreement between the total S1 charge collected by the PDS per muon in data (1830\,PE) and MC (1990\,PE), see Fig.~\ref{fig:S1Qtotal}. In a similar way, the observed light yield is computed for S1 signals acquired with the CRT trigger, see Table~\ref{tab:PEMeV}. In this case, based on the average track length in LAr observed in data (9.4\,m), an average deposited energy per muon of 1880\,MeV is obtained.
 
\begin{table}[!ht]
\begin{center}
\caption{Total S1 charge per cosmic muon (sum of all PMTs detecting a S1 signal) and observed light yield (average value from events and maximum value reached in an event) for two trigger modes. All the values obtained from data.}
\label{tab:PEMeV}
\begin{tabular}{ccccccc}
\hline
Trigger &  PMTs & S1 charge & \multicolumn{2}{c}{Obs. light yield} \\
& &  per muon (PE) &  \multicolumn{2}{c}{(PE/MeV)}\\
    \hline
 &    & & Avg. & Max. \\
    \hline
 & All & 1830 &2.3 & 21.4\\
Random & 6 TPB & 650  & 0.8 & 10.2\\
 &30 PEN & 1180 & 1.5 & 13.1\\
    \hline
& All & 1320 & 0.7 & 2.1\\
CRT & 6 TPB & 590 & 0.3 & 1.1 \\
&30 PEN & 730 & 0.4 &1.0 \\
    \hline
\end{tabular}
\end{center}

\end{table}

According to the values in Table~\ref{tab:PEMeV}, the complete PDS of ProtoDUNE-DP (36 PMTs) detects about 2.3\,PE/MeV and 0.7\,PE/MeV for random-trigger and CRT-trigger muon tracks, respectively. The latter case corresponds to an average track-PMT distance of 475\,cm whereas the former comprises muons crossing the detector at all distances from the PMTs (from close to distant tracks, up to 7\,m away from the PMTs). Since the scintillation photons from cosmic muons triggered with the CRTs are produced farther from the PMTs, the light is expected to undergo more attenuation. It is worth commenting that the difference between the average and maximum observed light yields is wider for random trigger (almost 10~times) than for CRT-trigger events (3~times) due to the larger variety of events (energies and track topologies) in the first case.

The light yield is also shown in Table~\ref{tab:PEMeV} separately for the two WLS groups in order to highlight the fact that the 6\,TPB PMTs alone collect 1/3 of the total light. It can be concluded that if the 36\,PMTs had TPB coating (PEN foil), a detection of about 5\,PE/MeV (2\,PE/MeV) would be reached in random trigger, on average. Therefore, in this particular PDS configuration, the use of PEN as the only WLS option would compromise the capability of the system as a calorimeter for low-energy particles. Considering TPB as the baseline WLS would make the low-energy physics goals of the DUNE Far Detector, such as triggering on a supernova neutrino burst, more feasible to accomplish. 




In conclusion, the better detection efficiency of the TPB-coated PMTs together with their stable performance during the detector operation confirm that TPB is a better WLS choice for future LArTPCs. However, the mechanical advantages of PEN make it a good candidate if the emission efficiency is not critical.



\section{Electroluminescence light detection}
\label{sec:s2}
\begin{figure*}[ht]
    \centering
    \includegraphics[width=0.85\textwidth]{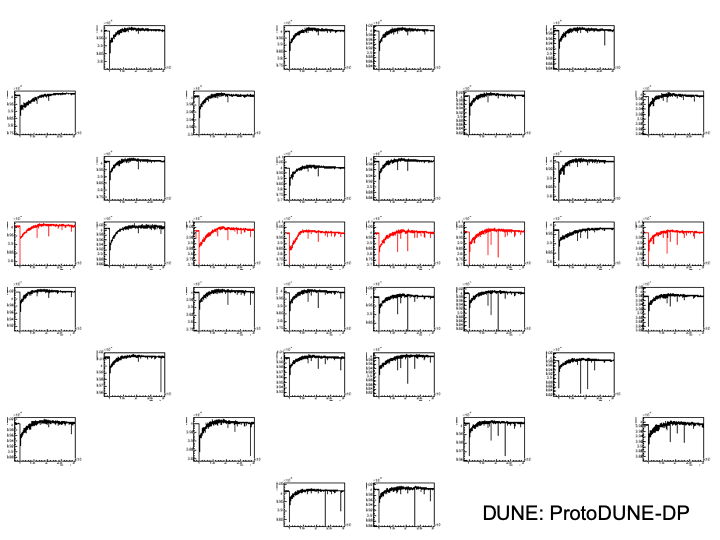}
    \caption{Example of an event with S1 and S2 signals showing the waveforms corresponding to the 36 PMTs according to their relative position in the detector. PEN PMTs are shown in black and TPB PMTs in red. The $x$-axis range is 0.8$-$3\,ms for all PMTs while the $y$-axis range varies for each PMT and is optimized to best display the  S2 signal. } 
    \label{fig:S2EventExample}
\end{figure*}

\begin{figure}[ht]
    \centering
    \includegraphics[width=0.46\textwidth]{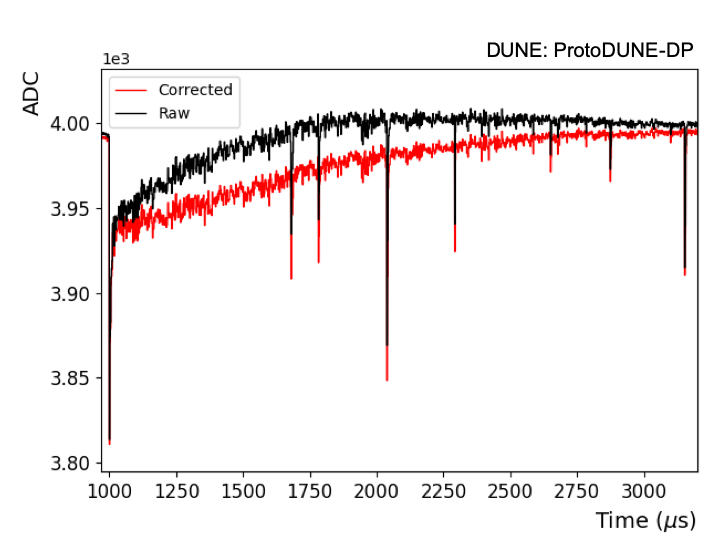}
    \caption{Example of a scintillation light event in a ProtoDUNE-DP PMT at a gain of $10^7$ in LAr with drift, amplification and extraction fields (S1 and S2 signal). The acquired waveform is shown in black and the resulting waveform after the overshooting correction in red.}
    \label{fig:S2WF}
\end{figure}

\begin{figure}[ht]
    \centering
    \includegraphics[width=0.46\textwidth]{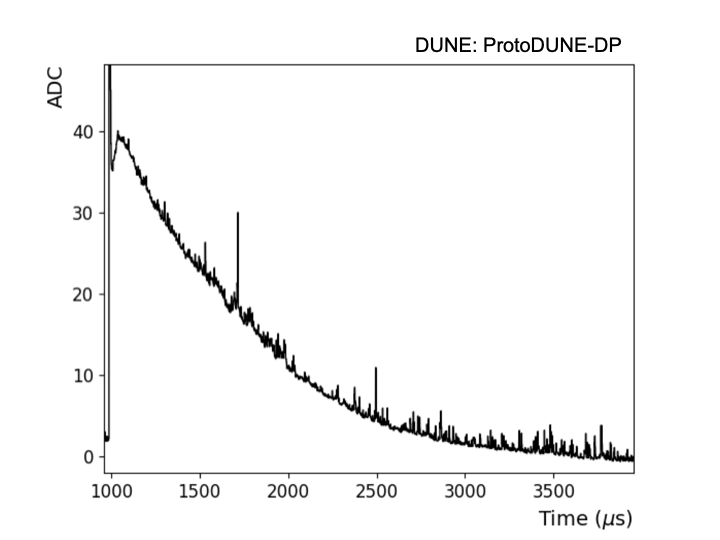}
    \caption{Average waveform of PEN PMTs with a S2 signal identified. The $y$-axis is zoomed to focus on the S2 signal.}
    \label{fig:S2Average}
\end{figure}

The electrons extracted into the gas phase produce produces a secondary scintillation signal, called S2, which is proportional to the drifted charge. The S2 signal provides information on the drifted electrons and the track topology. For instance, the time spread of the S2 signal is related to the track topology, as the more inclined the track is, the larger the S2 time spread is. The time difference between the S2 signal and the S1 peak corresponds to the drift time that takes to the electrons produced by the ionizing particles to the anode.

In ProtoDUNE-DP, S2 electroluminescence signals are detected in all PMTs when extraction and amplification fields are on, corresponding to light originating about 7\,m away from the PMTs. This detection is achieved thanks to the good LAr optical properties and purity and the high efficiency of the ProtoDUNE-DP PDS. In Fig.~\ref{fig:S2EventExample} an example event with S2 signal detected by all PMTs is shown. This is the first time light produced at such a large distance has been detected in a LArTPC.

ProtoDUNE-DP operated on the surface and observed a high S1 rate, see section~\ref{sec:muon}. The S2 signals are mostly observed as an increase in SPE rate which increases from hundreds of kHz, see section~\ref{sec:spe}, up to several MHz ($\sim$2.5\,MHz for TPB PMTs and $\sim$1.1\,MHz for PEN PMTs) when there is S2 light production (DP operation mode). It is clear that the S2 signals are an important and continuous contribution to the low-energy background in the PMT waveform when the extraction field is on.

As a result, not every S2 signal can be associated to its previous S1. Only very energetic events produce S2 signals that can be distinguished from the SPE background and associated to their previous S1. A dedicated algorithm was developed to select these events where the S2 signal can be distinguished from the SPE background and evaluate them in relation to their previous S1 signal.

An individual PMT waveform of a very energetic event selected by the algorithm can be seen in Fig.~\ref{fig:S2WF}. In this case, a fast S1 signal can be seen followed by a S2 signal with a duration of $\sim$2\,ms. Other S1 signals from cosmic muons are also visible. The S2 signals are observed to cause an overshoot in the waveforms as the amount of charge collected in the PMT anode exceeds the discharging rate of the combined PMT and readout circuit (1/$RC$ constant), effectively shifting the waveform baseline during the pulse. Waveforms are processed offline to correct for this effect, as described in~\cite{311light}. In Fig.~\ref{fig:S2WF} the PMT waveform before and after the overshooting correction is shown.

Fig.~\ref{fig:S2Average} shows the average waveform of the events passing the algorithm. These are very energetic events corresponding mainly to vertical showers where the track passed through the liquid-gas interface, and the S2 maximum is produced right after the S1. In these events the average S1 charge detected per PMT is $>$700\,PE and the S2 charge $>$30\,kPE. S2 signals last $\sim$2\,ms. As illustrated in Fig.~\ref{fig:mapfield}, vertical muons will traverse a field of $\sim$0.2\,kV/cm along the first two meters for which an electron drift velocity of $\sim$1\,mm/$\mu$s is expected~\cite{Lastoria:2748990}. This implies an expected S2 duration of $\sim$2\,ms, which is consistent with the observed time.





\section{Scintillation light in Xe-doped LAr}
\label{sec:xe}

The use of Xe-doped LAr is a promising alternative to pure LAr for large-scale LArTPCs, since it mitigates the light suppression due to some impurities and it also improves the detection efficiency and uniformity.



\begin{figure*}[!ht]
    \centering
    \includegraphics[width=\textwidth]{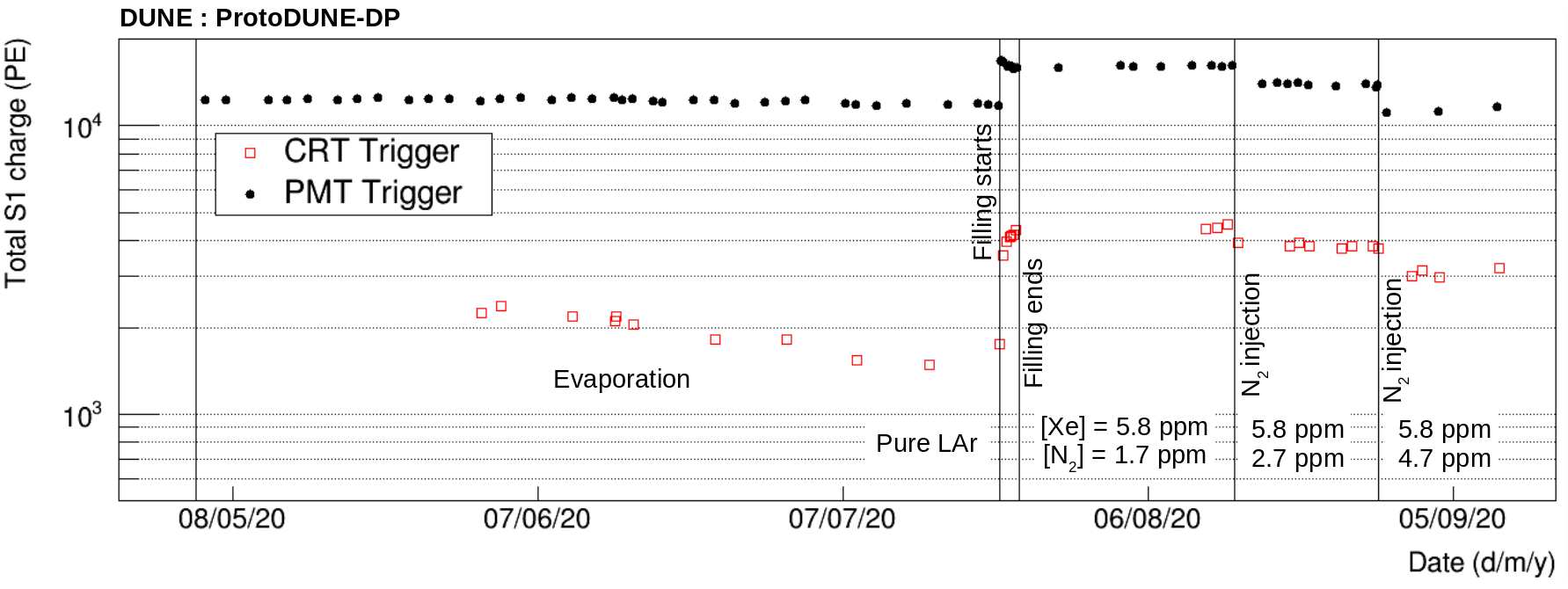}
    \caption{Evolution of the average S1 charge per event detected in all the PMTs for PMT(CRT)-trigger data in black (red) during the evaporation, filling and N$_{2}$ injections. The processes are described in section \ref{sec:data}. }
    \label{fig:Xenon_TotalLightYield}
\end{figure*}

In the presence of Xe, molecular Ar excimers in the triplet state live long enough to transfer their energy to the Xe atoms creating Xe excimers that decay and produce photons with a longer wavelength than 127\,nm~\cite{KUBOTA199371}. Therefore, the scintillation light is no longer monochromatic and has three components: Ar light at 127\,nm, and Xe light at 178\,nm and 150\,nm, with significant contribution at 150\,nm only at Xe concentrations below 1\,ppm~\cite{Neumeier_2015_ArXe}. Since molecular Ar excimers in the singlet state decay too fast to transfer their energy, they produce 127-nm photons only, while the late light is dominated by 150-nm and 178-nm photons.

Xe doping also affects the photon propagation. On one hand, it is reported that Xe acts as an impurity, suppressing part of the spectrum at 127\,nm even at the concentration of 0.1\,ppm~\cite{Neumeier_2015_Attenuation}. This absorption reduces the fast component of the detected signal. On the other hand, photons at longer wavelengths (150\, and 178\,nm) are not absorbed~\cite{Neumeier_2015_Attenuation}, and have a longer RSL. RSL is $\sim$1\,m for 127-nm photons while it is $\sim$3.5\,m for 150\,nm and $\sim$9\,m for 178\,nm photons~\cite{Babicz:2020den}. A longer RSL enhances the light detection at longer distances with respect to pure LAr, since photons with a longer RSL scatter less and are able to travel further thereby improving the uniformity of detection.

The presence of nitrogen in LAr leads to the suppression of the light production due to a quenching process driven by two-body collisions of $N_{2}$ impurities with excited argon excimers~\cite{Acciarri_N2}. This quenching affects mainly the excimers in the triplet state as the singlet state decays much faster. Since quenching by N$_{2}$ and Xe excimer formation are competing processes, the Xe atoms mitigate the light suppression due to the N$_{2}$. 

Additionally, the absorption of photons by N$_2$ is expected to increase with the concentration, with a reported absorption length for 127-nm photons of $28$ m at 1.7\,ppm, 20\,m at 2.7\,ppm and 12\,m at 4.7\,ppm~\cite{Jones:2013bca}. 

In the case of ProtoDUNE-DP, after the re-filling with $\sim$230 ton of Xe-doped liquid argon contaminated with N$_{2}$, as described in section~\ref{sec:data}, both N$_{2}$ and Xe species were present in the LAr, altering the light production and propagation. The effect of the presence of Xe and N$_2$ in the detected light in ProtoDUNE-DP is studied using two types of muon-track signals: First, events triggered with a TPB-coated PMT placed at the center of the detector with a minimum amplitude of 25\,PEs, for which the light is produced at a close distance from the PMTs, and second, CRT-trigger events, for which the PMT-track distance is in the range of 3$-$5\,m.

Fig.~\ref{fig:Xenon_TotalLightYield} shows the evolution of the total S1 charge detected by all PMTs as a function of time during the LAr evaporation from 7.4\,m to 5.1\,m of liquid level, re-filling and N$_{2}$ injection steps. Around 12\,kPEs are detected on average per PMT-trigger event, compared to only 2\,kPEs per CRT-trigger event in LAr. The S1 signal is integrated over a long time window of 12\,$\mu$s after the maximum.
The reduction of the collected light during the evaporation is visible for the CRT-trigger events, as part of the CRT-track is no longer in the liquid, while the collected light for the PMT-trigger tracks is stable. After the filling with Xe-doped LAr, the collected light increases as expected due to the longer RSL, and it decreases with the N$_{2}$ injections due to the quenching.



\begin{figure}[!ht]
    \centering
    \includegraphics[width=0.47\textwidth]{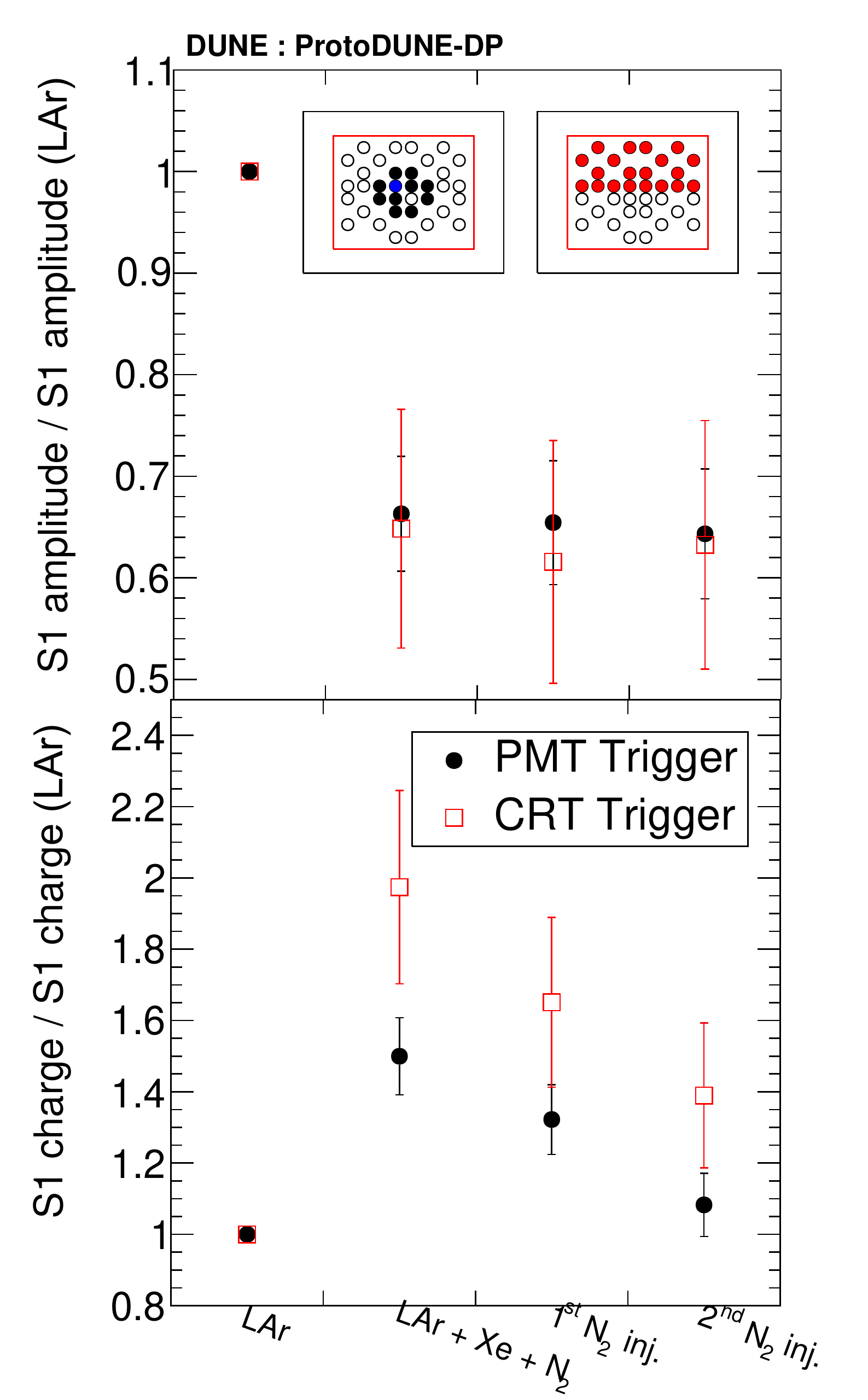}  
    \caption{Ratio of the average S1 amplitude and S1 charge in the three doping concentrations relative to pure LAr. PMT-trigger data are shown in black, and CRT-trigger in red. Only a selection of PMTs is considered for the average. Selected PMTs are marked as black (red) circles in the small diagram for the PMT(CRT)-trigger data. A blue circle marks the trigger PMT in the PMT-trigger diagram. Errors show the STD among the selected PMTs. Xe and N$_{2}$ concentrations for each situation are summarized in Table~\ref{tab:XenonData}.}
    \label{fig:Xenon_Filled}
\end{figure}

To quantify the impact of the presence of Xe and N$_{2}$ on the S1 charge and amplitude with respect to their values in pure LAr, data taken with the same liquid level are selected to ensure the same optical conditions. Fig.~\ref{fig:Xenon_Filled} shows the average variation of the S1 amplitude and S1 charge for different detector conditions. Only PMTs placed near the trigger PMT are considered for PMT-trigger data analysis, to have a similar track-PMT distance, while only channels placed below the CRT-track are selected for CRT-trigger data analysis, since the other PMTs are placed more than five meters away from the track and for these distant PMTs the signal is dominated by background. The amplitude decreases 35\% when adding 5.8\,ppm of Xe and 2.4\,ppm of N$_{2}$ with respect to pure LAr for both trigger modes, being unaffected by the N$_{2}$ addition. This reduction is due to the absorption of the 127-nm photons by the Xe atoms as reported in~\cite{Neumeier_2015_Attenuation}. Whilst a 16\% decrease of the fast component (S1 amplitude) for the CRT-trigger data is expected when adding N$_{2}$ according to~\cite{Jones:2013bca}, we observe the fast component remaining constant for the two N$_2$ injections. A possible explanation for this discrepancy is that the assumption of the fast component being monochromatic at 127\,nm is not accurate, and an additional light contribution at a different wavelength, that is not absorbed by the N$_{2}$, is masking the expected absorption at 127\,nm. The collected S1 charge increases 100\% for the CRT-trigger data, while only 50\% for the PMT-trigger data. This difference is understood as an improvement of the detection uniformity, since CRT-trigger muons are on average farther away from the PMTs, and the longer RSL of the Xe photons improves their collection at large distances. The decrease of the S1 charge due to the presence of N$_2$ is similar for both triggers (30\%), meaning that there is no dependence of the detected light suppression on the PMT-track distance. This indicates that the reduction is mainly due to the quenching of the Ar excimers by N$_{2}$ rather than photo-absorption.

\begin{figure}[!ht]
    \centering
    \includegraphics[width=0.5\textwidth]{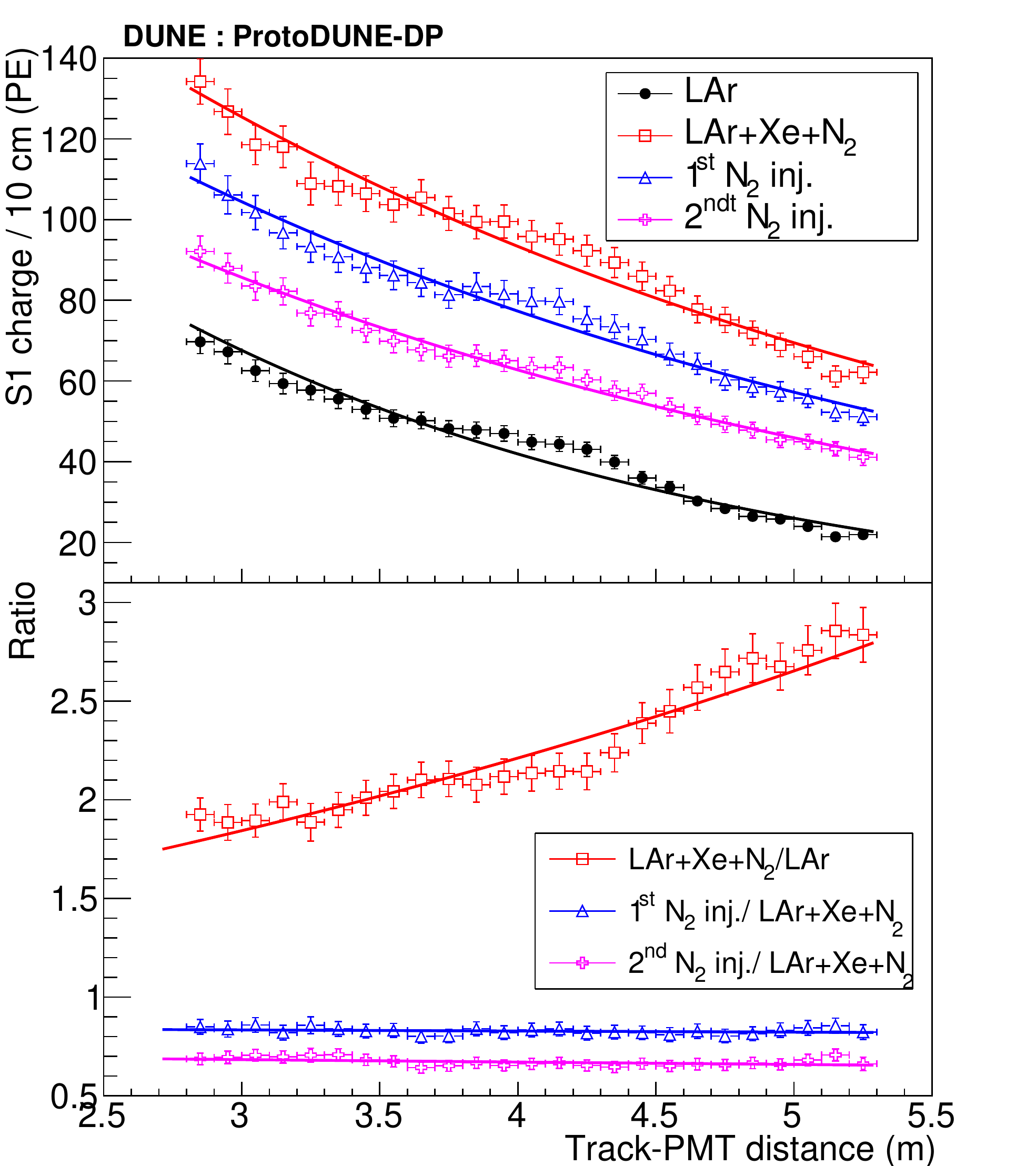}
    \caption{Top panel: Average collected S1 charge per PMT as a function of the track-PMT distance for the different doping concentrations. For each data-set, an exponential fit is performed (see results in Table~\ref{tab:XenonAttenuation}). Bottom panel: Ratios between the data shown in the top panel. Each data-set is fitted with an exponential function.}
    \label{fig:Xenon_ChargeVsDistance}
\end{figure}

In order to evaluate the effect of the Xe-doping on the attenuation length ($\lambda_\mathrm{att}$), a study of the dependence of the collected light per PMT with the track-PMT distance is performed for muon tracks crossing the CRT panels, see Fig.~\ref{fig:Xenon_ChargeVsDistance}. The range of distances is given by the position of the PMTs with respect to the triggered track, which is always approximately in the same position (see section~\ref{sec:intro}). The behavior is not purely exponential since light absorption by the field cage introduces a border effect. A shoulder shape is observed at around 4.3\,m, which corresponds to the PMTs placed at the center of the detector, where this effect is reduced. This limitation is due to the fixed geometry of the triggered tracks.
An exponential fit is performed to estimate the attenuation length in each detector condition, and the results are summarized in Table~\ref{tab:XenonAttenuation}. The effective attenuation length increases 60\% when adding Xe, as expected because of the longer RSL, and decreases 5\% when adding 2.9\,ppm of N$_{2}$. The lower panel of Fig.~\ref{fig:Xenon_ChargeVsDistance} shows the ratios of the top panel curves. The comparison between pure LAr and Xe-doped LAr in red shows the improved detection uniformity when adding Xe, with an increase of almost a factor of $\sim$3 on events at 5\,m and a factor of $\sim$2 on events at 3\,m. The extrapolation of this curve to short distances indicates an increase by a factor of 1.5 at 2.2\,m, as seen in Fig.~\ref{fig:Xenon_Filled} for the PMT-trigger muons, and no increase at 0\,m, as expected due to the longer RSL. The flat blue and magenta lines show that there is no dependence on the distance for the N$_{2}$ injections, as seen also in Fig.~\ref{fig:Xenon_Filled}.

\begin{table}[!ht]
\begin{center}
\caption{Measured attenuation lengths in the different doping concentrations from the exponential fits shown in Fig.~\ref{fig:Xenon_ChargeVsDistance}.}
\label{tab:XenonAttenuation}
\begin{tabular}{c c c c}
        \hline
    Situation & [Xe](ppm) & [N$_{2}$] (ppm)&  $\lambda_\mathrm{att}$ (cm) \\
        \hline
    LAr                   & 0   & 0   & $209\pm5$\\
    LAr + Xe + N$_{2}$    & 5.8 & 2.4 & $338\pm13$\\
    1$^{st}$ N$_{2}$ inj. & 5.8 & 3.4 & $332\pm13$\\
    2$^{nd}$ N$_{2}$ inj. & 5.8 & 5.3 & $321\pm12$\\
            \hline
\end{tabular}

\end{center}
\end{table}

\begin{figure}[!ht]
    \centering
    \includegraphics[width=0.5\textwidth]{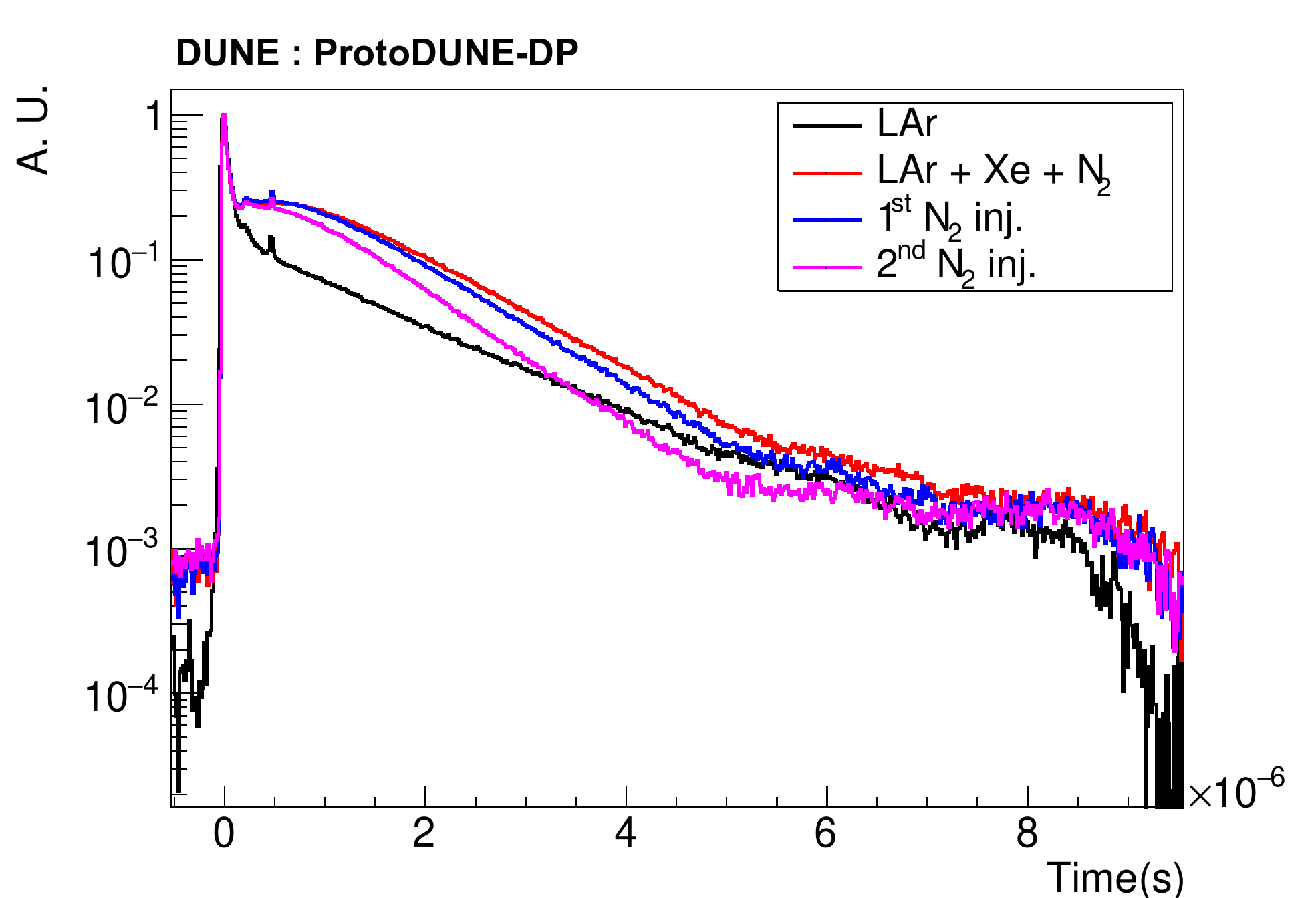}   
    \caption{PMT average waveform normalized to same amplitude in LAr (black), LAr + Xe + N$_{2}$ (red), after the 1$^{st}$ N$_{2}$ injection (blue) and after the 2$^{nd}$ N$_{2}$ injection (magenta). Concentrations are described in Table~\ref{tab:XenonAttenuation}.}
    \label{fig:Xenon_Profiles}
\end{figure}

The effect of xenon on the time profile of the waveforms is also studied. Fig.~\ref{fig:Xenon_Profiles} shows the average waveforms for a TPB-coated PMT in the different doping concentrations described in section~\ref{sec:data} (pure LAr in black), normalized to the same amplitude. The average waveform is the result of adding waveforms for a PMT placed near the trigger PMT. The second bump in the red, blue and magenta waveforms (when Xe is present) is the late light at 150\,nm and 178\,nm, as explained before. The profile of this second maximum changes with the N$_{2}$ injections, as the energy transfer rate from argon to xenon atoms is altered due to the quenching by N$_{2}$. The average waveforms of Xe-doped LAr are fitted to the sum of three exponential functions convolved with a Gaussian to account for the PMT response, in a similar way as in Eq.~\ref{eq:FitFunction}. In this case, the first exponential describes the fast signal, and the second and third exponential functions model the rise ($\tau_{\mathrm{transfer}}$) and decay ($\tau_{\mathrm{slow}}$) of the second bump from the Xe light.

The time constants obtained from the fits are shown in Table~\ref{tab:XenonTaus}. No difference between PEN-foil and TPB-coated PMTs is found.

\begin{table}[!ht]
\begin{center}
\caption{Time constants obtained from the fit of the average waveforms at the different doping concentrations. Errors show the variation among PMTs.}
\label{tab:XenonTaus}
\begin{tabular}{c c c}
        \hline
    Situation & $\tau_{\mathrm{transfer}}$ ($\pm0.01\mu$s) &  $\tau_{\mathrm{slow}}$ ($\pm0.01\mu$s) \\
        \hline
    LAr                        & N/A    & $1.43$ \\
    LAr + Xe + N$_{2}$         & $0.46$ & $1.10$\\
    1$^{st}$ N$_{2}$ injection & $0.41$ & $1.03$ \\
    2$^{nd}$ N$_{2}$ injection & $0.35$ & $0.91$ \\
            \hline
\end{tabular}

\end{center}
\end{table}

    
In summary, ProtoDUNE-DP data show that Xe doping is a promising technique for large-scale detectors like DUNE since it increases the collected light at large distances. With a small doping level of 5.8\,ppm of Xe (and even with the presence of 2.4\,ppm of N$_{2}$), an enhancement in the light detection efficiency (100\% increase for muons crossing at a distance of 3$-$5\,m from the PMTs) and a better uniformity (attenuation length 50\% longer) is measured. However, despite these advantages, it must be considered that the 35\% amplitude reduction observed in the fast signal 
could jeopardize the efficiency of a light-based trigger. 

\section{Conclusions}

ProtoDUNE-DP is a 6$\times$6$\times$6 m$^{3}$ LArTPC, operated at CERN between 2019 and 2020 to demonstrate the dual-phase technology at large scale for DUNE, a next generation long-baseline neutrino experiment. The photon detection system is composed of 36 8-inch cryogenic PMTs from Hamamatsu positioned at the bottom of the detector. The photon detection system collected cosmic-ray data for 18 months in stable conditions with all 36 PMTs in operation. The good performance validates the photon detection system design for future long drift distance LArTPCs.

ProtoDUNE-DP used PEN as a wavelength shifter for the first time in a large scale experiment and a comparison with the widely used TPB is carried out. TPB is estimated to be 3 times more efficient than PEN. The observed light yield from cosmic muons demonstrates that a system based exclusively on TPB as WLS would be needed to accomplish the DUNE low-energy physics program goals. Nonetheless, PEN can be taken into account as an alternative when the detection efficiency is not critical compared to the benefits of easy installation.



In ProtoDUNE-DP, considering the field limitations, it is found that at least 17\% of the scintillation light detected in the absence of a drift field comes from electron-ion recombination, verifying the expected trend from Birks’ law and in agreement with previous work. A decrease of $\tau_\mathrm{slow}$ with the drift field as reported in~\cite{311light} is also observed. An explanation of this effect is provided in~\cite{Segreto:2020qks} taking into account the quenching of the long lived triplet states through the self-interaction with other triplet states or through the interactions with molecular argon ions.


The size of ProtoDUNE-DP allows for an unprecedented study of the light propagation. An evaluation of the Rayleigh scattering length is carried out by comparing the measured light signals with the light predicted by the MC simulation testing two lengths (61.0\,cm and 99.9\,cm). The agreement between data and the 99.9-cm MC sample is better than for the shorter scattering length. It is also concluded that at least 11\% of the light detected by PMTs in the MC corresponds to VUV-light reflected off the field cage, cryostat walls, cathode and ground grid.

The cosmic muon flux in ProtoDUNE-DP at ground level is determined from the S1 signal rate detected by the PMTs and a cosmic-muon light simulation sample. The result, 148$^{+8}_{-11}$\,Hz/m$^2$, is consistent with other muon flux measurements in the literature \cite{microboone2021,mitrica2011}.  

The electroluminescence light, S2, produced in the gas phase about 7\,m away from the PMTs is observed in all 36 PMTs implying a high efficiency for the ProtoDUNE-DP photon detection system. The detected SPE rate increases up to several MHz when there is S2 light production, clear evidence that the S2 signals are an important and continuous contribution to the low-energy background in the PMT waveform.

Finally, ProtoDUNE-DP data has demonstrated the improvement of the light detection efficiency and uniformity in large LArTPCs, thanks to the Xe doping. A low doping level of 5.8\,ppm of Xe doubles the collected light at large distances (3$-$5\,m from the PMTs) even with the presence of 2.4\,ppm of N$_{2}$. However, it must be considered that the reduction observed in the fast signal amplitude could compromise the performance of a light-based trigger.

\begin{acknowledgements}




%
%
The ProtoDUNE-DP detector was constructed and operated on the CERN Neutrino Platform.
We gratefully acknowledge the support of the CERN management, and the
CERN EP, BE, TE, EN and IT Departments for NP04/Proto\-DUNE-SP.
%
%
This document was prepared by the DUNE collaboration using the
resources of the Fermi National Accelerator Laboratory 
(Fermilab), a U.S. Department of Energy, Office of Science, 
HEP User Facility. Fermilab is managed by Fermi Research Alliance, 
LLC (FRA), acting under Contract No. DE-AC02-07CH11359.
%
%
This work was supported by
CNPq,
FAPERJ,
FAPEG and 
FAPESP,                         Brazil;
CFI, 
IPP and 
NSERC,                          Canada;
CERN;
M\v{S}MT,                       Czech Republic;
ERDF, 
H2020-EU and 
MSCA,                           European Union;
CNRS/IN2P3 and
CEA,                            France;
INFN,                           Italy;
FCT,                            Portugal;
NRF,                            South Korea;
CAM, 
Fundaci\'{o}n ``La Caixa'',
Junta de Andaluc\'ia-FEDER,
MICINN, and
Xunta de Galicia,               Spain;
SERI and 
SNSF,                           Switzerland;
T\"UB\.ITAK,                    Turkey;
The Royal Society and 
UKRI/STFC,                      United Kingdom;
DOE and 
NSF,                            United States of America.
%
%
This research used resources of the 
National Energy Research Scientific Computing Center (NERSC), 
a U.S. Department of Energy Office of Science User Facility 
operated under Contract No. DE-AC02-05CH11231.
%

\end{acknowledgements}

\bibliographystyle{spphys}       

%
%
\bibliography{biblio}

\end{document}